\documentclass{article}
\usepackage{amsfonts}

\newcounter{lemmacounter}
\usepackage{graphicx}
\usepackage[dvips]{color}
\usepackage{amssymb}
\usepackage{pstricks}
\psset{unit=3cm}
\usepackage{url}
\usepackage[section]{placeins}	
\usepackage{floatrow}
\usepackage{caption}
\usepackage{alltt}
\usepackage{breqn}
\usepackage{listings}
\def\Euc{{\bf Euc}}

\def\B{{\bf B}}   
\def\R{{\mathbb R}}
\def\F{{\mathbb F}}

\def\implies{{\rightarrow}}

\newtheorem{Theorem}{Theorem}[section] 
\newtheorem{Proposition}[Theorem]{Proposition}

\def\PropNineFigure{%
\pspicture(3,2.6)(0,0.8)
\qline(0,1)(3,1)
\psdot(1,1)
\psdot(2,1)
\put(1,0.83){$A$}
\put(1.55,2.5){$B$}
\put(2,0.83){$C$}
\qline(1,1)(1.5,1.866)
\qline(2,1)(1.5,1.866)
\qline(1,1)(1.5,2.5)
\qline(2,1)(1.5,2.5)
\psline[linecolor=red](1.5,1)(1.5,2.5)
\psdot(1.5,1.866)
\psdot(1.5,2.5)
\put(1.53,1.866){$P$}
\endpspicture}

\def\InnerPaschFigure{%
\pspicture(-0.4,0)(1.8,1.8)
\psdot(0,0.6)
\put(0,0.47){$a$}
\pscircle(1,0.6){0.03}
\put(0.97,0.47){$x$}
\qline(0,0.6)(0.97,0.6)
\psdot(1.33,0.6)
\put(1.38,0.55){$q$}
\qline(1.03,0.6)(1.33,0.6)
\psdot(1.2,1.4)
\put(1.2,1.45){$c$}
\qline(0,0.6)(1.2,1.4)  
\qline(1.2,1.4)(1.375,0.3) 
\psdot(1.375,0.3)
\put(1.43,0.25){$b$}
\psdot(0.54,0.96)
\put(0.43,1.0){$p$}
\qline(1.375,0.3)(1.015,0.584)  
\qline(0.54,0.96)(0.979,0.62)  
\endpspicture}

\def\OuterPaschFigure{%
\pspicture(1.8,1.8)
\psdot(0,0.6)
\put(0,0.47){$b$}
\psdot(1,0.6)
\put(1,0.47){$p$}
\qline(1.2,1.4)(1.375,0.3) 
\pscircle[fillcolor=white,fillstyle=solid](1.33,0.6){0.03}
\put(1.38,0.55){$x$}
\qline(0,0.6)(1.3,0.6)  
\psdot(1.2,1.4)
\put(1.2,1.45){$q$}
\qline(0,0.6)(1.2,1.4)  
\psdot(1.375,0.3)
\put(1.43,0.25){$a$}
\psdot(0.54,0.96)
\put(0.43,1.0){$c$}
\qline(1.375,0.3)(1.015,0.584)  
\qline(0.54,0.96)(1.375,0.3)  
\endpspicture}

\def\TarskiFiveSegmentFigure{%
\psset{unit=2.25cm}
\pspicture(2.5,0)(2.2, 1.3)
\qline(0.0,0.15)(2.0,0.15)
\qline(0.0,0.15)(1.0,0.85)
\psline[linestyle=dashed](1.0,0.85)(2.0,0.15)
\qline(1.0,0.85)(0.7,0.15)
\put(1,0.9){$d$}
\put(0.0,0){$a$}
\put(0.7,0){$b$}
\put(2.0,0){$c$}
\qline(2.5,0.15)(4.5,0.15)
\qline(2.5,0.15)(3.5,0.85)
\psline[linestyle=dashed](3.5,0.85)(4.5,0.15)
\qline(3.5,0.85)(3.2,0.15)
\put(3.5,0.9){$D$}
\put(2.5,0){$A$}
\put(3.2,0){$B$}
\put(4.5,0){$C$}
\psset{unit=3cm}
\endpspicture}

\def\EuclidParallelRawFigure{%
\pspicture(3.2, 1.1)
\pspolygon[fillstyle=solid,fillcolor=yellow](0.67,0.7)(0.78,0.425)(1.49,0.7)
\pspolygon[fillstyle=solid,fillcolor=yellow](0.07,0.15)(0.78,0.425)(0.9,0.15)
\qline(-0.15,0.15)(2.47,0.15)
\qline(2.53,0.15)(2.7,0.15)
\qline(-0.15,0.7)(2.7,0.7)
\qline(0.0,0.9)(2.473,0.163)
\qline(2.5275,0.14)(2.7,0.085)
\pscircle(2.5,0.15){0.03}
\psdot(0.67,0.7)
\put(0.67,0.78){$p$}
\psdot(1.29,0.515)
\put(1.28,0.39) {$a$}
\psdot(0.9,0.15)
\put(0.88,0.04) {$q$}
\psdot(0.07,0.15)
\put(0.07,0.04){$s$}
\qline(0.67,0.7)(0.9,0.15)
\psdot(1.49,0.7)
\put(1.5,0.78) {$r$}
\qline(1.49,0.7)(0.07,0.15)  
\psdot(0.78,0.425)
\put(0.7,0.3){$t$}
\qline(0.9,0.15)(1.49,0.7)  
\put(-0.3,0.12) {$L$}
\put(-0.3,0.68) {$K$}
\put(-0.15,0.88) {$M$}
\endpspicture}

\def\SameSideFigure{%
\pspicture(0,0.4)(1.7,1.6)
\qline(0.7,1)(2,1)
\put(0.6,0.97){$p$}
\put(2.05,0.97){$q$}
\psdot(0.9,0.5)
\put(0.83,0.35){$a$}
\psdot(1.2,0.5)
\put(1.22,0.35){$b$}
\psline(1.8,1.5)(1.2,0.5)
\psline(1.8,1.5)(0.9,0.5)
\pscircle[fillstyle=solid,fillcolor=white](1.5,1){0.03}
\put(1.54,0.87){$y$}
\pscircle[fillstyle=solid,fillcolor=white](1.346,1){0.03}
\put(1.27,1.05){$x$}
\pscircle[fillstyle=solid,fillcolor=white](1.8,1.5){0.03}
\put(1.85,1.5){$c$}
\endpspicture
}

\def\OppositeSideFigure{%
\pspicture(0.7,0.4)(1.7,1.6)
\qline(0.7,1)(2,1)
\put(0.6,0.97){$p$}
\put(2.05,0.97){$q$}
\psdot(0.9,0.5)
\put(0.83,0.37){$a$}
\psdot(1.8,1.5)
\put(1.85,1.5){$b$}
\psline(1.8,1.5)(0.9,0.5)
\pscircle[fillstyle=solid,fillcolor=white](1.346,1){0.03}
\put(1.27,1.05){$x$}
\endpspicture
}

\def\ETPasteFigure{%
\psset{unit=2cm}
\newrgbcolor{lightblue}{0.8 0.8 1}
\pspicture(4.5,1)(0,-0.2)
\pspolygon[fillcolor=yellow,fillstyle=solid](0,0.5)(2,0)(2,0.5)(0,0.5)
\pspolygon[fillcolor=yellow,fillstyle=solid](2.5,0.5)(3.5,-0.17)(4,0.5)(2.5,0.5)
\pspolygon[fillcolor=lightblue,fillstyle=solid](0,0.5)(1.5,1)(2,0.5)(0,0.5)
\pspolygon[fillcolor=lightblue,fillstyle=solid](2.5,0.5)(3.5,1.17)(4.0,0.5)(2.5,0.5)

\qline(0,0.5)(2,0.5)
\qline(0,0.5)(2,0)
\qline(0,0.5)(1.5,1)
\psdot(0,0.5)
\psdot(2,0.5)
\psdot(2,0)
\psdot(1.5,1)
\psdot(2.5,0.5)
\psdot(3.5,1.17)
\psdot(3.5,-0.17)
\put(-0.2,0.4){$A$}
\put(2.05,0.45){$D$}
\put(1.56,0.97){$B$}
\put(2.05,-0.05){$C$}
\put(2.3,0.4){$a$}
\put(4.05,0.45){$d$}
\put(3.55,1.18){$b$}
\put(3.56,-0.2){$c$}
\psset{unit=3cm}
\endpspicture}

\title{Proof-checking Euclid}        
\author{Michael Beeson \\Julien Narboux \\ Freek Wiedijk}        
\date{\today
}  

\begin{document}
\maketitle

\begin{abstract}We used computer proof-checking methods to verify the 
correctness of our proofs of the propositions in Euclid Book I.  
We used axioms as close as possible to those of Euclid,  in a language 
closely related to that used in Tarski's formal geometry.  We used
proofs as close as possible to those given by Euclid,  but filling
Euclid's gaps and correcting errors.  Euclid Book I has 48 propositions;
we proved 235 theorems.  The extras were partly ``Book Zero'',  preliminaries
of a very fundamental nature,  partly propositions that 
Euclid omitted but were used implicitly, partly advanced theorems that 
we found necessary to fill Euclid's gaps,  and partly just variants of Euclid's
propositions.  We wrote these proofs in a simple fragment of first-order logic
corresponding to Euclid's logic,  debugged them using a custom software tool, and 
  then   checked them in the well-known and trusted proof checkers
HOL Light and Coq. 
\end{abstract}

\section{Introduction}  Euclid was the ``gold standard'' of rigor
for millenia.  The {\em Elements} of Euclid set the standard of proof used by Isaac Newton in his 
{\em Principia}  and even Abraham Lincoln claimed to have read all ten books
of Euclid and learned from it how to prove something in court.  The 
{\em Elements} also inspired the form of the American Declaration of Independence.  In the modern era,  beginning already in the nineteenth 
century,  the standards of proof in mathematics became more demanding,
and the imprecisions and gaps in Euclid were more apparent than before.
Even before that time, some mathematicians focused on the perceived flaw that 
the Fifth Postulate  (the ``parallel postulate'', or ``Euclid 5'')  was less intuitively evident
than the other four.%
\footnote{Proclus \cite{proclus}, writing in the 
fifth century, said that Euclid 5 needed a proof, and tried
to supply one;  and he was not the first, since he also criticized a previous attempt by Ptolemy.
See \cite{greenberg}, Ch.~5 for further history.}
 Efforts to remove this ``flaw'' by proving Euclid 5 
led to the development of non-Euclidean geometry, showing that in fact 
Euclid 5 was a necessary postulate, not a flaw.  

Nevertheless there are flaws in Euclid, and we want to discuss their
nature by way of introduction to the subject.%
\footnote{It is customary to 
refer to the propositions of Euclid with notation like I.44, which means
Proposition 44 of Book I.''}
The first gap occurs in the first proposition, I.1,  in which Euclid 
proves the existence of an equilateral triangle with a given side, by 
constructing the third vertex as the intersection point of two circles.
But why do those two circles intersect?  Euclid cites neither an axiom
nor a postulate nor a common notion.
This gap is filled by adding the ``circle--circle'' axiom, according to 
which if circle $C$ has a point inside circle $K$,  and also a point
outside circle $K$, then there is a point lying on both $C$ and $K$.

There is, however, a second gap in the proof of I.1.  Namely, the 
proposition claims the existence of a {\em triangle},  which by definition
is (or at least is determined by)  three {\em non-collinear} points.%
\footnote{Euclid never defines {\em triangle}, although he does define
{\em right triangle} and {\em equilateral triangle}.  Instead he 
mentions {\em trilateral figure}, which is ``contained by three straight lines.''  We read that to imply non-collinearity, for otherwise there 
would be only one line, and nothing ``contained.''} 
Why are the three points in question not all on a line?  Intuitively,
if they were, then one would be between the other two, violating Euclid's
common notion that the part is not equal to the whole, since all the sides
are equal.  

A formal proof of Prop.~I.1 cannot follow Euclid in ignoring this 
issue.  The proposition that if $B$ is between $A$ and $C$ then 
$AB$ is not equal to $AC$ therefore must precede proposition I.1,  unless
one is to consider it as an axiom formalizing one of Euclid's common notions.
In the next section, we discuss the axioms and postulates of Euclid, 
and how we have chosen to render them in modern first-order form.

The two gaps in I.1 illustrate two common failures.   Many of the gaps 
in Euclid fall into one of these categories:
\medskip

(i) A failure to prove that a point clearly 
shown in the diagram actually exists, e.g. that two lines really do 
intersect, or as in I.1  two circles.
\smallskip

(ii) A failure to prove that points shown in the diagram to 
be non-collinear,  are in fact non-collinear.
\smallskip

(iii) A failure to prove that a point shown in the diagram to 
be between two other points,  is in fact between those points.
\medskip

Another example of an error of type (i) is in the second paragraph of Prop.~I.44,
``let $FG$ be drawn through to $H$''.  Here $H$ has not been proved to exist,
a strange omission in that a few lines later Euclid does feel the need to use 
Postulate 5 to prove that $K$ exists; but then two lines later ``let $HA$, $GB$
be produced to the points $L$, $M$''.  That is, the lines shown as intersecting at $L$
and $M$ do in fact intersect--but Euclid offers no justification for that line of 
the proof.  There are dozens of such fillable gaps in Euclid's proofs, some more 
easily filled than others. 

Not every error in Euclid falls into these categories, however.  
Consider Prop.~I.9,  in which Euclid bisects an angle.  The method 
is to lay off the same length on both sides of the angle, and then
construct an equilateral triangle on the base thus formed.  Connecting
the vertex of the original angle with the vertex of the equilateral 
triangle, we get the angle bisector. 
\begin{figure}[ht]
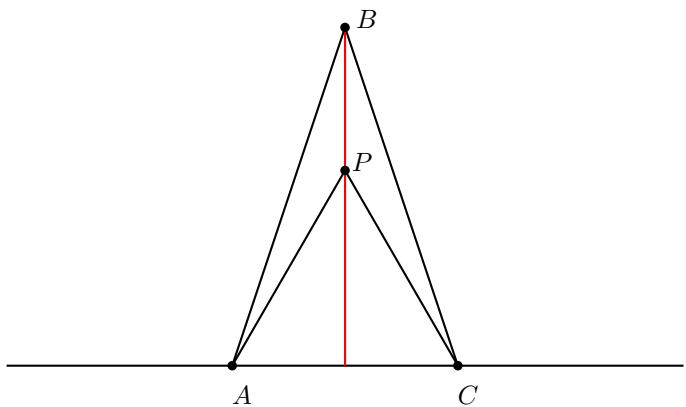

\PropNineFigure
\caption{Euclid's proof of I.9.  What if $B=P$?  Even if $B \neq P$,
why does $BP$ lie in the interior of $ABC$, i.e., why does it meet $AC$?}
\end{figure}
 Oops, but the two points might 
coincide!   Well,  in that case we ought to have drawn the {\em other}
equilateral triangle, on the other side of the base.  But Prop.~I.1
did not provide for the construction of two triangles, and we cannot 
easily construct ``the other one.''  We certainly need to expand the list 
of ways in which Euclid's proofs fall short by at least two more items:
\medskip

(iv) A failure to prove that points shown in the diagram to be distinct are in fact distinct.
\medskip

(v)  A failure to show that points are on the same side (or opposite sides) of a 
line, as the diagram shows them to be.
\medskip

Even if we could solve these problems, the proof of I.9 still would not be correct,
since we would still need to show that the angle bisector constructed does in fact 
lie in the interior of the angle.  
That difficulty brings us to an important point.  There is no 
``dimension axiom'' in Euclid to guarantee that we are doing plane 
geometry.  Hence ``circles'' are really ``spheres'', and rather than 
just two equilateral triangles on a given base, there are infinitely 
many.  $BP$ might not even lie in the same plane as $AC$!
So even if the vertex of the equilateral triangle is distinct
from the vertex of the original angle,  why does the line between them 
lie in the interior of the angle?  In fact that would be a problem even 
in two dimensions--an example of (iii) above.  

Even if it were possible to fix that problem by adding 
a dimension axiom,  that would not be desirable.  Euclid didn't just 
forget to write down a dimension axiom.    In Book X and beyond, 
Euclid works in three-space, and the culmination of the whole series 
of books is the study of the Platonic solids.  Hence it is clear that
Euclid is not meant to be restricted to plane geometry. 
In the absence of a dimension axiom, it is good advice to the reader
to visualize ``circles'' as spheres.   Then the two circles used in I.1 have 
not just two but many intersection points.   The circle--circle axiom only 
guarantees the existence of {\em one} intersection point.%
\footnote{Since Euclid did not 
state the axiom, one could consider strengthening it to state the existence of an
intersection point on a given side of the line connecting the centers of the 
circles.  But Euclid also did not say in I.1 or I.9 anything about this problem; 
and we {\em are} able to prove I.9 from the more fundamental form of circle--circle.}
Therefore we conclude that Euclid's proof of I.9 is fatally flawed.  We prove 
it another way:  by first showing how to bisect a line, and then using that to 
bisect an angle.  The simple and ingenious proof that a line can be bisected if 
it is the base of an isosceles triangle was apparently not noticed until Gupta's 
thesis in 1965 \cite{gupta1965},  but could very well have been discovered by Euclid.

The fact that there is no dimension axiom is not always remembered in Book~I,
and is the source of several difficulties. 
 A good example is Prop.~I.7,  which 
says that if $ABC$ is a triangle and $ABD$ is another triangle 
congruent to $ABC$ (that is, $AC=AD$ and $BC=BD$), and $C$ and $D$ 
are on the same side of $AB$, then $C$ is equal to $D$.   Look at the 
figure for I.7;  in two dimensions it appears contradictory,  but as soon as 
you think that it might be in three dimensions,  the contradiction disappears.
The statement 
of Prop.~7 does correctly include the hypothesis that $C$ and $D$ are 
on the same side of $AB$, but Euclid never uses that hypothesis in the 
proof.  That is not surprising, since he never defined ``same side'', 
so he had no possible way to use that hypothesis.  It should have been 
used to verify the claim that angle $DCB$ is less than angle $DCA$, because
there is a point of intersection of $AD$ and $CB$.  This turns out to 
not be provable, even after proving a number of more basic propositions 
in ``Book Zero'';  we could not prove I.7 without using ``dropped
perpendiculars'',  which are only constructed much later in Euclid.

A well-known geometer told us ``there are no errors in Euclid'', in the 
sense that the statements of all the propositions are true in the 
plane.  If we supply Tarski's definition of ``same side'',  an even 
stronger version of that claim is true:  the statements of all the propositions
are true in every finite-dimensional space $\R^n$.  However, the same 
cannot be said of the {\em proofs}.  Many of these have problems
like those of I.9 and I.7; that is, we could fix these problems only 
by proving some other propositions first, and the propositions of 
the first half of Book I had to be proved in a different order, 
namely 1,3,15,5,4,10,12,7,6,8,9,11, and in some cases the proofs are 
much more difficult than Euclid thought.  After proving those early propositions, we could follow Euclid's order better, and things went well 
until Prop.~44.  In Propositions~44 and 47 there are numerous points of 
difficulty, which took extra propositions to resolve.  As one example 
we mention the proposition that every square is a parallelogram, which 
Euclid uses implicitly in proving Prop.~47. (By definition a square has
equal sides and four right angles, and a parallelogram is a quadrilateral
with opposite sides parallel.)  Euclid could and should have proved that.

The aims of the project reported on in this paper are as follows:
\medskip

(i) Fix Euclid's axioms (and common notions), using an axiom system rather 
close to Euclid's,  but including axioms about betweenness that Euclid omitted,
and with other changes discussed below.
\smallskip

(ii) Give correct proofs of all the propositions in Book I from the 
new axioms, following Euclid's proofs as closely as possible.
\smallskip

(iii) Show that those proofs are indeed correct by checking the proofs
using the proof-checking programs HOL Light and Coq.    
\smallskip

(iv)  Show that the axioms are indeed correct by computer-checking proofs
that the axioms hold 
in the Cartesian plane $\R^2$.%
\footnote{The axioms hold in $\F^n$,
  where $\F$ is any Euclidean field (an ordered field in 
which positive elements have square roots), but we have not 
computer-checked a proof of that fact, even for $n=2$; we did check $\R^2$.}  
\medskip

Accordingly,  in this paper we limit the discussion of geometry to the 
description of the axioms,  the description of a few specific flaws in Euclid's
reasoning, and a discussion of Euclid's notion of ``equal figures'' that is 
necessary to verify the axioms we use about that notion. 
Our focus in this paper is on proof checking.   What we report on here would 
still have been worth doing, even if there were no gaps or errors in Euclid.
 The details of our proofs, 
and a discussion of the errors in the original proofs of Euclid,  will be published separately,
with a focus on the geometry and on the correspondence between the axioms
and proofs of Euclid and those of our formal development.  That will 
necessitate a longer discussion than is possible here.

The formal proofs themselves, as well as the PHP and ML scripts that we used,
are available on the Web.  Look for links to them at 
\smallskip

\noindent
\url{http://www.michaelbeeson.com/research/CheckEuclid/index.php}

\noindent
They are also available as ancillary files to the version of this paper
posted on ArXiv.

\section{History}

A scholarly history of the previous attempts to axiomatize Euclidean geometry would 
require a long paper in itself; we offer only a few highlights here. But 
before beginning, we point out that axiomatizing Euclidean geometry is not 
quite the same as axiomatizing Euclid:  the former aims at an axiom system
that permits the derivation of important theorems, regardless of whether the 
axioms or the proofs are similar to Euclid's axioms and proofs; the latter
pays attention to those points.
 
The best known attempt is Hilbert's 1899 book \cite{hilbert1899}.  Hilbert
had been a vocal proponent of the axiomatic method, and his book was probably 
meant partly to illustrate that method on the example of geometry.   First-order
logic was in its infancy and Hilbert's system was not first order.  He made use of 
Archimedes's axiom, and his continuity ``axiom'' is a strange mixture of logic and model theory.%
\footnote{It can be viewed as a second-order axiom involving quantification over
models of the other axioms.}
The fundamental idea to use betweenness and congruence as primitive relations goes back 
to Pasch\cite{pasch1882}.   Further contributions by Mollerup \cite{mollerup1904}, Veronese \cite{veronese1891},
Pasch \cite{pasch1882}, and Peano\cite{peano1889} are discussed below in connection 
with the axioms they helped to develop.  After Hilbert, the most important work is
the axiom system of Tarski.  This is a first-order system,  and not only is it first-order,
it is points-only, meaning that there are variables only for points.   Lines are given
by two points,  and angles by three points, and equality between angles is a defined notion.
This fits Euclid very well:  Euclid almost always refers to lines by two points,  and angles
by three points.   Tarski's system was developed in 1927,  but publication was long delayed;
for the history see the introduction to (the Ishi press edition of) \cite{schwabhauser}. 
Although a development of Euclid in Tarski's system could have been done in the 1960s at Berkeley,
it was not done. Instead efforts focused on reducing the number of Tarski's axioms by 
finding dependencies, and on proving fundamental results like the existence of perpendiculars
and midpoints without using the parallel postulate or any continuity axioms. 
The results of these efforts finally appeared in \cite{schwabhauser},  which contains the remarkable results of \cite{gupta1965}. 

In spite of a century of effort, in 2017 
we did not possess any formal analysis of Euclid's proofs, for 
Hilbert and Tarski had both aimed at avoiding the circle axioms and 
developing segment arithmetic,  while Euclid uses the circle axioms 
and the last half of Book I is based on the equal-figure axioms (discussed 
in \S\ref{section:figures}). 
Even if we add the circle axioms, the last half of Book~I can be proved
in Hilbert or Tarski's theories only after the long and difficult development 
of segment arithmetic, so that ``equal figures'' can be defined as ``equal area.'' 

It is now 35 years since the publication of \cite{schwabhauser}, and  meantime, the 
technology of proof checking by computer has advanced. 
 Our predecessors stopped at the threshold, so to speak, by 
working on neutral geometry and minimal axioms systems, instead of formalizing Euclid. 
We serendipitously find ourselves in the situation where 
it is possible for us to take up that task, and also to verify 
(using existing computer proof checkers)  that our proofs are flawlessly correct.  

\section{Language}

Euclid did understand the fundamental point 
that not every fact can be proved;  the first fact accepted could not be 
proved because there would be nothing to prove it from.  But he 
did not understand that similarly, not every concept can be defined.
Thus he famously attempted to define ``point'' and ``line'' and 
``rectilinear angle.''  (The Greeks also considered ``angles'' formed by 
curved sides.)   These natural-language ``definitions''  are not
mathematical usable; so in practice Euclid treated points, angles, 
circles, and lines all as primitive notions.  In addition, Euclid treated
``figures'' as a primitive notion, in the sense that he never made
use of the circular and vague definitions he offered.%
\footnote{ A figure is that which is 
contained by a boundary; a boundary is that which is an extremity 
of anything.}
In Book I,  only triangle and quadrilaterals are used.
   Euclid also accepted concepts
of ``equal'' and ``greater than'' for each of points, angles, circles, lines,
and figures without 
definition (but curiously, there is no ``greater than'' for figures).
In the middle of the nineteenth century,  it was 
recognized that ``betweenness'' and ``equidistance'' were good primitives
for geometry,  and later it was realized that it is possible to work with angles 
represented by triples of points, instead of taking them as primitive, 
so angle equality and inequality are defined concepts.  That is what we 
do in our formal work.  Thus all our axioms, except those mentioning circles,
 are formulated in a 
``points-only'' language, in which the fundamental relations are 
betweenness and equidistance.  

Betweenness is a 3-ary relation $\B(A,B,C)$, which Euclid
wrote as ``$B$ lies on the finite straight line $AC$,'', or 
(for example in Prop.~I.14) as ``$AB$ and $BC$ lie in a straight line.'' 
 We interpret $\B$ as {\em strict} betweenness, i.e. the endpoints do 
not lie on the line.%
\footnote{Hilbert used strict betweenness; Tarski used non-strict 
betweenness, on purpose because the degenerate cases could be used
to reduce the number of axioms. We use strict betweenness, on purpose
to avoid degenerate cases that express unintended things and have to 
be separately worried about.}

Collinearity is the relation $L(A,B,C)$ expressing that either two 
of the points are equal or one lies between the other two.  This is 
a statement about points only.   It seems that 
for Euclid, lines were primitive objects,  rather than sets, and
the incidence relation (point lies on line) too fundamental even to 
notice, as it does not occur in the list where ``point'' and ``line'' 
are ``defined.''  Whether close to Euclid or not, we use the first-order
formulations of betweenness (a primitive) and collinearity (defined).%
\footnote{Most of the time Euclid's lines are finite, which 
may confuse a modern reader at first, since today finite lines are 
called ``segments'', and ``line'' means ``infinite line.''
Hilbert made (infinite) lines primitive objects, but treated 
finite lines (segments) as sets of points. 
}

 ``Equidistance'' is a 4-ary relation 
representing the congruence of finite lines, ``$AB$ is equal to $CD$''.
Euclid, or at least his translator Heath,
 used ``equal'' rather than ``congruent''. (The word ``equidistance''
is also not faithful to Euclid, who never spoke of distances.) 
  
 There is one exception to our ``points-only'' approach.  
In order to follow Euclid more closely, we allow giving a name to a 
circle.  Circles are given by point and radius, so we can say 
``$J$ is a circle with center $P$ and radius $AB$.''  (Here the
``radius'' is not a number but a finite line.)  This is expressed
by the formula $CI(X,C,A,B)$, 
using a primitive relation symbol $CI$.
We do not make use of equality between
variables of the sort for circles. 

Euclid never gives circles a single-character name as we do.
  Nor does he name circles by center and radius,  or center
and point-on-circle.   Instead he names circles by listing three points
that lie on the circle.  One of those points is a point that appears
to exist in the diagram, and is conjured into existence without proof 
by the act of naming it.  This naming technique papers over the 
lack of the circle--circle axiom in Euclid, and introduces a gap into the proof 
every time it is used.  We therefore {\em must} deviate from Euclid's naming 
convention for circles.  

Betweenness and equidistance are sufficient as primitive relations for 
elementary geometry,
but the latter part of Euclid Book I  uses another primitive relation,  ``equal figures'',
which is discussed in \S\ref{section:figures}.  We mention it here only to note that 
the complete definition of our language requires inclusion of the primitive relations 
discussed in \S\ref{section:figures}.

\section{Definitions}\label{section:definitions}
\begin{figure}[ht]
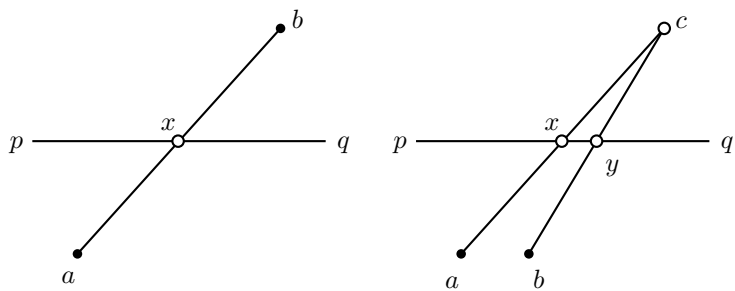

\center{\OppositeSideFigure \SameSideFigure}
\caption{(Left) $a$ and $b$ are on the opposite side of $pq$.
(Right) $a$ and $b$ are on the same side of $pq$ if there exist
points $x$ and $y$ collinear with $pq$, and a point $c$,
 such that $\B(a,x,c)$ and $\B(b,y,c)$.}
\label{figure:sides}
\end{figure}

Euclid gives a long list of definitions at the beginning of Book~I.
We do the same.  Euclid's list has some important omissions, notably
``same side'' and ``opposite side''.   These are defined in Fig.~\ref{figure:sides}.%
\footnote{This definition is due to Tarski \cite{schwabhauser}.  Hilbert had planes
as a primitive concept, and discussed ``same side'' and ``opposite side'' only in 
the context of a fixed plane, using a definition that would not work without having 
planes as a primitive concept.  Tarski's definition of ``same side'' is vital for 
making possible a points-only formalization that would work in more than two dimensions.
}  
Euclid defined ``square'' but omitted ``parallelogram'' and ``rectangle''. 
He defined ``parallel lines'' to be lines that do not meet but lie in the 
same plane (thus illustrating that his omission of a dimension axiom was
no accident!)  On the other hand, he failed to define ``lies in the same 
plane''.   Once we have defined ``same side'' as in Fig.~\ref{figure:sides},
it is easy to define ``lies in the same plane'', as each line and point 
not on the line determine two half-planes, together making up a plane.
In the formal statement of ``same side'',  a line is specified by
two distinct points $p,q$, and the incidence of $x$ on that line is 
expressed by 
``$p,q,x$ are collinear''.   Formally we use the relation $L(p,q,x)$ defined
above in terms of betweenness.   This definition exemplifies how one works with 
points only, avoiding the explicit mention of lines.  The price one has to pay for 
this simplification is that one then has to prove that it doesn't matter which 
particular points $p,q$ we chose to represent the line.  That is,  ``same side''
is invariant if $p$ and $q$ are changed to some other pair of distinct points 
each of which is collinear with $p$ and $q$.  

Euclid's failure to define ``lies in the same plane'' leaves us to 
complete his definition of ``parallel''.  First, we discuss the ``not meeting''
part of the definition.  Lines in Euclid are finite, but ``parallel'' is 
about infinite lines.  So ``$AB$ does not meet $CD$'' means that no matter 
how those two finite lines are produced, the lengthened lines still do not 
have a point in common.  In other words, there is no point collinear with 
both $AB$ and $CD$.  On the other hand,  ``$AB$ crosses $CD$'' means
that there is a point both between $A$ and $B$, and between $C$ and $D$.

 We define ``Tarski-parallel''
by ``$AB$ and $CD$ do not meet, and $C$ and $D$ lie on the same side of 
$AB$.''   This is clearly not what Euclid intended, as to Euclid it 
seems obvious that if $AB$ is parallel to $CD$ then $CD$ is parallel to 
$AB$.  So we define instead that $AB$ is parallel to $CD$ if there is no 
point collinear with both $AB$ and $CD$,  and there are 
four points $a,b,c,d$ with $a$ and $b$ collinear with $AB$, and $c,d$
collinear with $CD$, and $ad$ crosses $bc$.  With this definition, there
is a very short proof of the symmetry property.  On the other hand, 
the two definitions can be proved equivalent.  It follows that if $AB$ 
and $CD$ are parallel then $A$ and $B$ are on the same side of $CD$, which 
is quite often actually necessary,  but never remarked by Euclid. 

Euclid defines a square to be a quadrilateral with at least one
 right angle, in which 
all the sides are equal.%
\footnote{But in I.46 and I.47 the proofs work as if the definition 
required all four angles to be right, so we take that as the definition.}
He does not specify that all four vertices lie in 
the same plane.  This is not trivial to prove, but we did prove it, so Euclid's
definition does not require modification.   Euclid does not define 
``rectangle''.  One would like to define it as a quadrilateral with four right 
angles.  It is a theorem that such a figure must lie
in a plane.  However,  the proofs we found involve reasoning ``in three dimensions''.
Even though Euclid Book I has no dimension axiom, and we must therefore 
be careful not to assume one, nevertheless all the {\em proofs} in Book~I 
deal with planar configurations.   We therefore define ``rectangle'' to be a quadrilateral with four 
right angles, whose diagonals cross,  that is, meet in a point.  This condition 
is one way of specifying that a rectangle lies in a plane.  We can then prove 
that a rectangle is a parallelogram.  

The Appendix contains a complete machine-generated list of our definitions.

\section{Angles} 
We take only points and circles as primitive objects.  Angles are 
treated as ordered triples of non-collinear points, $ABC$.   The 
point $B$ is the vertex of the angle.  Equality of angles is 
a 6-ary relation, which we write informally as ``angle $ABC = abc$''.
The definition is that there exist four points (one on each side of 
each angle) that form, with the vertices $B$ and $b$, 
two congruent triangles.   (Two triangles are 
congruent, by definition, if all three pairs of corresponding sides are 
equal.)%
\footnote{This is not the same definition as used in \cite{schwabhauser}, 
but it works, and seems simpler to us;  perhaps the one in \cite{schwabhauser} 
seems simpler in the presence of function symbols for line extension.}  
This definition does not permit ``straight angles'',  ``zero angles'', 
or angles ``greater than 180 degrees.''  Such ``angles'' are also not 
used in Euclid. 

A point $F$ {\em lies in the interior of angle $ABC$} if it is 
between two points lying on the two sides of the angle.  Angle 
ordering is defined by $abc < ABC$ if angle $abc$ is equal to angle 
$ABF$, for some $F$ in the interior of $ABC$.   Note that these 
definitions make sense without any dimension axiom; that is, they 
work fine in three-space.  

We then have to prove as theorems those properties of angle equality 
and ordering that Euclid assumed as ``common notions'':  reflexivity, 
symmetry, and transitivity of angle equality;  the fact that 
angle $ABC$ is equal to angle $CBA$; transitivity of angle ordering. 
The fact that an angle cannot be both equal to and less than the 
same angle is quite difficult to prove,  although taken for 
granted by Euclid in several proofs.  That is, of course, the key 
result needed to prove antisymmetry and trichotomy for angle ordering.

Hilbert \cite{hilbert1899}  took angles as primitive, and had an 
axiom about copying angles that specified the {\em uniqueness} of 
the copied angle.   The uniqueness assumption builds in as an axiom 
the property that an angle cannot be both less than and equal to itself.
Since this can be proved, it might be considered an imperfection to assume it 
as an axiom.   While Hilbert took angles and equality of angles as 
primitive, he did define angle ordering just as we do.  Because of 
the uniqueness part of his angle-copying axiom, he had no difficulty 
proving trichotomy.

\section{Axioms and Postulates}
Euclid had three groups of what would now be called axioms:  common notions,
axioms, and postulates.   The common notions were intended to be principles of 
reasoning that applied more generally than just to geometry.  For example, what 
we would now call equality axioms.   The axioms and postulates were about geometry.
The distinction between an ``axiom'' and a ``postulate'', according 
to Proclus \cite{proclus}, p.~157, is that a postulate asserts that some point can be constructed,
while an axiom does not.  In modern terms an ``axiom'' is purely universal,
while a postulate has an existential quantifier.

Heath's translation lists five common notions, five postulates, and zero
axioms.  Simson's translation \cite{simson} lists three postulates, twelve axioms,
and zero common notions.  The extra axioms are discussed by Heath on p.~223 of \cite{euclid1956},
where they are rejected.

\subsection{Euclid's Common Notions}

 Euclid's first common notion is ``things equal to the same thing are equal to each other.''
That is, 
$$a=c \land b=c \implies a=b.$$
Modern mathematicians would prefer
$$a=c \land c=b \implies a=b.$$
But then, they need symmetry as a separate axiom ($a=b \implies b=a$),  while 
that can be proved from Euclid's axiom above.  We follow Euclid in this matter, 
although of course it is of no serious consequence.%
\footnote{Apparently Euclid considered symmetry too
obvious to mention.
Or maybe, he considered ``$A=B$'' and ``$B=A$'' to just be
different expressions of the same proposition, rather than different 
but equivalent propositions.}

Euclid's fourth common notion is ``Things which coincide with one 
another are equal to one another.'' We take this to justify
reflexivity. We consider the transitivity, reflexivity, and symmetry
of point equality to be part of logic.

When ``equality'' refers to congruence of 
lines, these principles correspond to three congruence axioms. 
We also need the
axiom that $AB$ is equal to $BA$. In other words, Euclid's lines are not directional. Euclid never explicitly states
this principle, but it is often necessary when formalizing his proofs.
Perhaps Euclid would regard $AB$ and $BA$ as 
``coincident'', in which case this axiom is covered by common
notion 4, quoted above.

Euclid himself used common notion 4  to justify his ``proof'' of 
the SAS principle I.3 by ``superposition''.  We reject this proof,
and that leaves only $AB=BA$ and reflexivity of congruence, point
equality, and figures to correspond to common notion 4. 

Angle equality
is a defined concept and its properties are theorems, not axioms.
Equality of figures is axiomatized in \S\ref{section:figures}.

Euclid's fifth common notion is ``The whole is greater than the part.''
In our formalism, inequality of finite lines $AB < CD$ is defined as 
``$AB$ is equal to $CE$
 for some $E$ between $C$ and $D$.''  
Thus common notion 5 (for lines) is built into the 
 definition.   Then ``the whole is not equal to 
 the part'' (for lines) becomes $AB < CD$ implies $AB$ is not equal to $CD$,  
 which boils down to the principle {\tt partnotequalwhole}:
 $$ \B(A,B,C) \implies \neg\, AB=AC.$$
This is a theorem, not an axiom, in our development.

Euclid's second common notion is ``If equals be added to equals,
the wholes are equal'', and the third common notion is ``If equals be
subtracted from equals, the remainders are equal.''
Common notion 2 becomes our axiom 
{\tt sumofparts}, which says that if $AB=ab$ and $BC=bc$ and 
 $\B(A,B,C)$ and $\B(a,b,c)$, then $AC = ac$.  Here $AB$ and $BC$ 
 are ``parts'' and $AC$ is the ``whole'' made by ``adding'' the two lines.
 The related principle {\tt differenceofparts}, corresponding to
common notion 3, is proved, rather than assumed as an axiom; there is a fuller
discussion in \S\ref{section:degenerate}.  Angle inequality is a 
defined notion and there are no axioms about it.  These two common 
notions justify several of the axioms for figures, which are discussed
in \S\ref{section:figures}.

Equality also enjoys the substitution property for each predicate in our language:
\begin{eqnarray*}
\B(a,b,c) \land a=A \land b=B \land c=C \implies \B(A,B,C)   \\
ab=cd \land a=A \land b=B \land c=B \land d=D \implies AB=CD  
\end{eqnarray*}
and similarly for the predicates for ``equal figures.'' 
In practice, the proofs are checked assuming the second-order property:
$$ a=b \implies \forall P\,( P(a) \iff P(b)) $$
which allows to avoid introducing a separate axiom for each predicate.
That is, we do not actually use the substitution axioms for individual predicates,
but allow the substitution of $A$ for $B$ in any derived formula, when $A=B$ or $B=A$
has been derived.  Such substitutions are often needed in formalizing 
Euclid's proofs, so even if he did not explicitly state the principle,
he understood it.

A complete machine-generated list of the common notions is in the Appendix. 

\subsection{Betweenness Axioms}
Euclid never made explicit mention of betweenness, ignored all places
where it should have been proved,  and had no axioms for proving 
betweenness statements.  We will not discuss the historical origins
of the following axioms, nor the possibilities for reducing their number
(this is certainly not a minimal set, but the proofs required to eliminate
some of these axioms are long and difficult.)  We give them the names
they are given in our formal development, which is why there are no spaces
in those names.

\begin{eqnarray*}
\neg \B(a,b,a)  && \mbox{\tt betweennessidentity} \\
\B(a,b,c) \implies \B(c,b,a) && \mbox{\tt betweennesssymmetry} \\
\B(a,b,d) \land \B(b,c,d) \implies \B(a,b,c) &&\mbox{\tt innertransitivity} \\
\end{eqnarray*}

The following axiom is called {\tt connectivity}  and can be rendered
in English as ``If $B$ and $C$ lie on the finite straight line $AD$,  and neither is between $A$ and the other,  then they are equal.''

Formally:
$$
\B(a,b,d) \land \B(a,c,d) \land \neg\, \B(a,b,c) \land \neg\, \B(a,c,b) \implies b=c
$$

This principle was expressed in antiquity as ``a straight line cannot enclose an area.''%
\footnote{ It is discussed by Proclus \cite{proclus}, p.~126,  who thinks it superfluous
as it is included in the meaning of Postulate 1.  Apparently Simson was not convinced, 
as his translation \cite{simson}
lists it as an axiom.  Heath rejects it as an axiom (p.~232 of \cite{euclid1956}),  not on mathematical 
grounds, but because he came to the conclusion that it is an ``interpolation'',  i.e., is not 
in the original Euclid,  in spite of being included in three of the ``best manuscripts.''
}
It is closely related to the principle known in modern times as ``outer connectivity'',
which says that if line $AB$ has two extensions $C$ and $D$ then either $C$ is between
$B$ and $D$ or $D$ is between $B$ and $C$.  We prove outer connectivity 
as a theorem from the connectivity axiom.%
\footnote{Outer connectivity was discussed already by Proclus, who stated it as 
``two straight lines cannot have a common segment'' \cite{proclus}, p.~168-9, \S216-17.
Proclus says it is implicit in Euclid's line extension axiom. Neverthless, Proclus considers some 
possible proofs of it--but not the ingenious proof offered by Potts in the commentary to Prop.~I.11
in \cite{potts1845}, p.~14, which shows that outer connectivity follows from perpendiculars and the 
fact that an angle cannot be less than itself.   The latter, however, is a difficult theorem,
if it (or a close equivalent) is not assumed as an axiom. 
}

\subsection{Extension of lines}
Euclid postulated that every line can be extended, but (at least in Heath's translation)
 did not 
say by how much.%
\footnote{The Simson translation \cite{simson} renders the extension postulate as 
{\em ``That a terminated straight line may be produced to any length in a straight line.''}
Perhaps Euclid's extension postulate said more than Heath's translation indicates.
}
We render Euclid 1 as 
\medskip

\centerline{If $A \neq B$ then there exists $C$ with $\B(A,B,C)$.}
\medskip

Tarski postulated instead that every line $AB$ 
can be extended by the amount $CD$;  that is, there exists 
a point $E$ such that $\B(A,B,E) \land BE = CD$.  Since Euclid's
lines have distinct endpoints, it should be required that both $A \neq B$ and 
$C \neq D$.

There is an intermediate form we call {\tt localextension}, in which 
you are allowed to extend $AB$ by the amount $BC$; that is, the 
segment used to measure the extension and the segment to be extended
have a common endpoint.  

Euclid's Prop.~I.2 asserts that 
given any point $A$ and line $CD$ there is a point $E$
with $AE = CD$.    
The Tarski extension postulate renders Prop.~I.2 superfluous.  
That is a matter of some 
regret, since I.2 has a beautiful proof.  Clearly the Tarski extension
postulate goes beyond what Euclid had in mind; therefore we assume 
only Euclid 1 as stated above.

The line-circle axiom says that if $P$ is inside circle $C$ then 
any line through $P$ meets $C$.  This axiom is discussed fully in 
\S\ref{section:circleaxioms} below, but it is relevant to 
line extensions, because if we have a circle with center at $B$,
then line-circle enables us to extend $AB$ by the radius of the circle.
In this way one shows that Euclid 2 and line-circle together imply
{\tt localextension}, since the circle needed for a local extension can be 
drawn with a collapsible compass.  With {\tt localextension}, we 
can carry out the proof of I.2,  and after I.2 we can prove the Tarski
extension principle.  From there on the development is unaffected
by the choice not to assume the Tarski extension principle as an axiom,
but our weaker axiom exactly reflects what Euclid used and permits us 
to prove I.2 as he did.  

The key to this development is the ``bootstrapping'' aspect of it:
one needs I.2 to prove the Tarski extension principle with line-circle,
but one needs some extensions by a given amount to prove I.2, so it 
appears at first that Euclid 1 is not sufficient; so we first prove
{\tt localextension}, then use it to prove I.2, then prove Tarski
extension. For this to work it is also necessary that ``inside'' 
be correctly defined and that the line-circle axiom be correctly formulated;
there are several wrong ways to do these things. 
Euclid did not give us a complete proof to follow, since he did not state or use
any version of line-circle, and did not define ``inside'' which is the
key task in formulating line-circle.

\subsection{Five-line Axiom}
Euclid attempted, in Proposition I.4,  to prove the side-angle-side
criterion for angle congruence (SAS).  But his ``proof'' appeals to 
the invariance of triangles under rigid motions, about which there is 
nothing in his axioms, so for centuries it has been recognized that 
in effect SAS is an axiom, not a theorem.

\begin{figure}[ht]
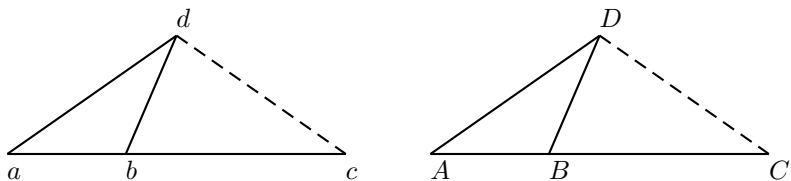

\center{\TarskiFiveSegmentFigure}
\caption{If the four solid lines on the left are equal to the 
corresponding solid lines on the right, then the dashed lines
are also equal.}
\label{figure:5-segment}
\end{figure}

Instead of SAS itself, we take an axiom known as the ``five-line
axiom.''  This axiom is illustrated in Fig.~\ref{figure:5-segment}.
 Its conclusion is, in effect, the congruence of 
triangles $dbc$ and $DBC$ in that figure. 
  Its hypothesis expresses the congruence (equality, in Euclid's
phrase) of angles $dbc$
and $DBC$ by means of the congruence of the exterior triangles $abd$ and 
$ABD$.   
\FloatBarrier
Our version of the five-line axiom was introduced by Tarski,
although we have changed non-strict betweenness to strict betweenness.%
\footnote{The history of this axiom is as follows.
The key idea (replacing reasoning about angles by reasoning
about congruence of segments) was
introduced (in 1904) by J. Mollerup \cite{mollerup1904}.
His system has an axiom closely related to the 5-line axiom,
and easily proved equivalent.  Tarski's version \cite{tarski-givant}, however, is 
slightly simpler in formulation.   Mollerup (without comment)
gives a reference to 
Veronese \cite{veronese1891}.  Veronese does have a theorem 
(on page 241) with the same diagram as the 5-line axiom, and
closely related,  but he does not suggest an axiom related to this 
diagram.}

\subsection{Pasch's Axiom}

\begin{figure}[ht]
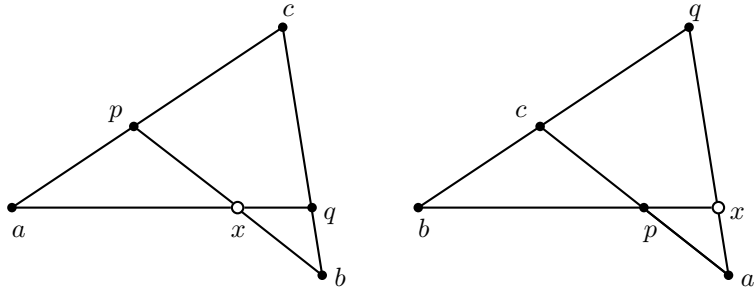

\center{\InnerPaschFigure\OuterPaschFigure}
\caption{Inner Pasch (left) and outer Pasch (right).  Line $pb$ meets triangle $acq$ in one side $ac$, and meets an extension of side $cq$.  Then 
it also meets the third side $aq$.
 The open circles show the points asserted to exist. }
 \label{figure:pasch}
\end{figure}

Pasch \cite{pasch1882} introduced the axiom that bears his name, 
in the form that says that if a line enters a triangle through one side, 
it must exit through another side (or vertex).  That version, of course, 
is only true in a plane.  Seven years later, Peano \cite{peano1889} 
introduced what are now called ``inner Pasch'' and ``outer Pasch'', 
which work without a dimension axiom.%
\footnote{Axiom XIII in \cite{peano1889} is outer Pasch, with $\B(a,b,c)$
written as $b \in ac$.  
Axiom XIV is inner Pasch.  Peano wrote everything in formal symbols 
only, and eventually bought his own printing press to print his books himself.}
 See Fig.~\ref{figure:pasch}.
In that figure, we use the convention that solid dots indicate points 
assumed to exist, while an open circle indicates a point that is asserted 
to exist.

Technically ``Pasch's axiom'' should be ``Pasch's postulate'', since
it makes an existential assertion,  but the terminology is too 
well-established to change now.

\subsection{Degenerate cases.} \label{section:degenerate}
 Tarski was always interested in 
minimizing the number of axioms, and was happy if allowing ``degenerate''
cases in axioms allowed one to combine what would otherwise be several
separate axioms.  Tarski's versions of Pasch's axiom 
allow all the points to lie on one line.   From those degenerate cases
one can derive basic principles about the order of points on a line
that we take as separate axioms.  Our reason for requiring a 
non-collinearity hypothesis in inner and out Pasch is the principle
that axioms should correspond to intuition.  If you have to draw a 
different picture to convince yourself that the degenerate case is valid,
then that is a different intuition; so generally it should be a different 
axiom.   

Another kind of ``degenerate case'' is the so-called ``null segment''
$AA$.   
In Euclid, 
lines are given by two {\em distinct} points, so there is no 
such thing as the ``line'' (or segment) $AA$.  Formally we have the 
predicate $E(A,B,C,D)$.  What happens then when $A=B$?  The 
idea that $E(A,B,C,D)$ means that line $AB$ is equal (congruent) to 
line $CD$ suggests that it should be false when $A=B$, since line $AA$
does not exist.  But the idea that $E(A,B,C,D)$ means ``equidistance''
suggests that $E(A,A,C,D)$ should be equivalent to $C=D$.  Are null segments allowed, or not?  If they are,  we need the axiom $E(A,A,C,C)$ that
says ``all null segments are equal''.   If they are not, we need 
$\neg E(A,B,C,C)$.  Euclid gives us no guidance: he only works with
lines that have distinct endpoints (no null segments), but he never 
says a single word about null segments.  We were (eventually)
able to follow Euclid in that respect:  we have no axioms either way
about null segments, and our axiom system has models with null segments
and models without null segments.  We are ``agnostic'' about null segments.

 It is not a
fundamental philosophical issue, as the talk of ``null segments'' is 
just shorthand; nobody suggests that $AA$ is really a line.  The 
argument against null segments is that we want to follow Euclid closely.
The argument for null segments is that
allowing $E(A,A,C,C)$ is occasionally convenient in allowing the succinct
statement of theorems.  For example, congruence is preserved under
reflection in a point, which is called {\tt pointreflectionisometry}.
 That is stated by saying that if $B$ is 
the midpoint of both $AC$ and $PQ$, then $AP=CQ$.  That statement 
includes the case when $A=P$:  then the conclusion is $Q=C$.  To keep
our theory agnostic, we had to assume $A \neq B$ in  {\tt pointreflectionisometry}, which complicated the proof of 
{\tt linereflectionisometry}. (These theorems do not occur in Euclid 
anyway, but we need them to prove Euclid's Postulate 4, all right angles
are equal.) 

The picture of the 5-line axiom illustrates that it is meant to 
capture the intuition leading to the SAS congruence principle.
But when the picture collapses onto a line, the 5-line axiom makes
``one-dimensional'' or ``linear'' 
assertions about order and congruence of points
on a line. 
The ``degenerate cases'' of the 5-line axiom are 
when point $D$ lies on the line $ABC$.
 What principles are embodied in 
those degenerate cases?  The cases $D=A$ is
{\tt sumofparts}:  if $B(A,B,C)$ and $B(a,b,c)$, and $AB=ab$ and $BC=bc$,
then $AC=ac$.  The case $D=B$ in the 5-line axiom similarly is 
{\tt differenceofparts}: if $B(A,B,C)$ and $B(a,b,c)$, and $AB=ab$ and $AC=ac$,
then $BC=bc$.  Euclid did not state these principles explicitly, but 
when he used them, he referred to his common notions about ``adding equals
to equals'' and ``subtracting equals from equals''. 

Following the principle that there should be a one-to-one correspondence
between intuitions (or diagrams) and axioms, we should impose the extra
hypotheses in the 5-line axiom that $D$ is not collinear with $AC$ 
and $d$ is not collinear with $ac$.  If one does that,  one will 
need to add at least four more ``linear'' axioms (that are now proved
with the help of the 5-line axiom) One needs more than 
{\tt sumofparts} and {\tt differenceofparts}, and we did not 
discover an ideal set of axioms to add.  Moreover, one can argue
intuitively for the unmodified axiom as follows:  when the 
fourth point is not on the line, the intuition for the axiom is SAS.
But now think of the fourth point moving onto the line; if it approaches
a limit on the line,  all the quantities mentioned vary continously,
so the congruences in the 5-line axiom remain true in the limit.  The 
unmodified axiom makes proofs using it insensitive to the distinction
whether point $D$ is or is not on line $AC$, and hence it supports
intuitionistic proofs,  which a restricted version would not.  
For these reasons, we retained the unrestricted 5-line axiom.

\subsection{Euclid's Postulate 5}
\label{section:euclid5}

Euclid's ``parallel postulate'',  or ``Euclid 5'',  is a postulate rather
than an axiom, because it asserts that two lines meet, i.e.,  there 
exists a point on both lines.  The hypothesis as Euclid stated the 
postulate involves angles.  We use instead a ``points-only'' version.
Then Euclid's version becomes a theorem.  

\begin{figure}[ht]
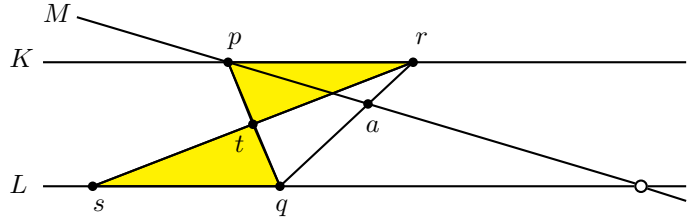

\caption{Euclid~5.  Transversal $pq$ of lines $M$ and $L$ makes corresponding interior angles less than 
two right angles, as witnessed by $a$. The shaded triangles are assumed congruent. Then $M$ 
meets $L$ as indicated by the open circle.}
\label{figure:EuclidParallelRawFigure}
\hskip 2.5cm
\EuclidParallelRawFigure
\end{figure}

Most modern geometry textbooks replace Euclid 5 by ``Playfair's axiom''
(introduced by Playfair in 1729), which asserts the uniqueness of a 
line parallel to a given line $AB$, through a point $P$ not collinear with $AB$.
This also becomes a theorem in our development.  Although it does not 
occur as a proposition in Euclid,  it is several times used 
implicitly in Euclid's proofs.%
\footnote{
Explicitly:  Playfair is used directly in propositions 44, 45, 47,
and indirectly in 37, 38, 42, 46; and more indirectly in 39, 40, 41, 42, 48;
so overall it is used in 39-48 except 43.  Euclid 5 is used directly in 29, 39, 42, 44 
and indirectly in 29-48 except 31, which is the existence of the parallel line.  
Euclid should have proved 31 before 29, to emphasize that Euclid 5 is not needed for it.
}

\subsection{Euclid's Postulate 4}
Euclid 4 says ``all right angles are equal.''  
The definition of a right angle is this:  $ABC$ is a right angle 
if there is a point $D$ such that $\B(A,B,D)$ and $AB = DB$ and $AC=DC$.  
It has been claimed since the time of Proclus that Euclid~4 is provable,
but since the axioms and definitions were not so precise, we are not 
certain that any of the alleged proofs could be counted as correct 
until the proof in Tarski's system presented in \cite{schwabhauser}. 
In our system this is a difficult proof, depending on the fact that 
both reflection in a point and reflection in a line are isometries 
(preserve congruence and betweenness).  The proof has to work without 
a dimension axiom.  It is a very beautiful proof and obviously much 
deeper than the ``proofs'' given by Proclus and Hilbert.  The beautiful
part of the proof (after the observation that reflections are isometries)
is contained in Satz~10.15 of \cite{schwabhauser}.   

Even though this proof is difficult, it would clearly be a flaw to assume Euclid 4 as an 
axiom,  when it can in fact be proved.  Therefore we prove it, 
rather than assume it.%
\footnote{One may well ask, if we find it necessary to prove Euclid 4  ``just because we can'',
why do we not find it necessary to prove one of the two Pasch axioms, inner and outer Pasch,
from the other ``just because we can''?   The answer is that we still need one of them 
as an axiom, and the same intuition that justifies one of them also justifies the other.  
Therefore there is no {\em conceptual} economy in reducing the number of axioms by one.
But proving Euclid~4 does offer a conceptual simplification.}

\subsection{Circle construction axioms}
To express our theory 
in first-order predicate calculus, we use
a two-sorted predicate calculus, one sort for points and one for circles.
$CI(J,A,B,C)$ means that circle $J$ is a circle with
 center $A$ and radius $BC$.

Euclid's Postulate 3 is ``To describe a circle with any centre and distance.''
By this, he meant that you can draw a circle with a given center and passing
through a given point.  This is often called the ``collapsible compass'' 
construction, as opposed to the ``movable compass'' or ``rigid compass'',
that permits drawing
a circle with given center and radius specified by a given line 
(which need not have the center as an endpoint).  Euclid's Prop.~I.2 
shows that the collapsible compass can imitate the movable compass.  Past formal
systems could not capture the difference.  But in our system, 
Euclid~3 is directly rendered in our language as 
$$ A \neq B \implies \exists J\, CI(J,A,A,B)$$
while the movable compass is 
$$ B \neq C \implies \exists J\, CI(J,A,B,C).$$

But {\em a priori}, $J$ might also have center $P$ and radius $QR$. 
That this is not the case is the content of Euclid's Prop.~III.1.
The statement of III.1 is ``To find the centre of a given circle'',
but the proof proceeds by showing that two supposedly different centers
must in fact coincide.  In our formalism that is stated:
if $CI(J,A,B,C)$ and $CI(J,a,b,c)$ then $A=a$.  (It can be proved
that the radii are equal too: $BC=bc$.)

We define $on(P,J)$ to mean that for some $A$, $B$, and $C$,
$J$ is a circle with center $A$ and radius $BC$, and $AP=BC$. 
We do not want a circle to be just a triple of points.
That a circle is determined by its center
and radius is expressed by saying that if $J$ is a circle with 
center $A$ and radius $BC$, and $P$ is on $J$, then $AP=BC$. This
axiom is called {\tt circle-center-radius}; see the listing of axioms
in an Appendix for a formal statement.

The predicates
``inside'' and ``outside'' can be defined using inequality of finite lines, 
and the circle--circle continuity axiom can be translated straightforwardly
from the informal English version given above.   We then define
relations ``on'' and ``inside'' and ``outside''  that take both 
a point argument and a circle argument.

\subsection{Circle continuity axioms}
\label{section:circleaxioms}
Euclid had no postulates or axioms about circles other than Euclid 3.  
There are three continuity axioms in the 
literature:
\smallskip

{\em Circle--circle}:  if circle $C$ has one point inside circle $K$
and one point outside, then there is a point on both circles.
\smallskip

{\em Line--circle}:  if line $L$ has a point $P$ inside circle $K$, then 
there are two points $A$ and $B$ on both $L$ and $K$, such that 
$P$ is between $A$ and $B$.  
\smallskip

{\em Segment--circle}: if line $L$ has a point $A$ inside circle $K$ and 
a point $B$ outside, then there is a point on $K$ between $A$ and $B$.
\smallskip

The reader should bear in mind that in the absence of any dimension axioms,
a ``circle'' is ``really a sphere'', or even some kind of ``hyper-sphere''.  

Circle--circle is used twice in Book I,  once in Prop.~I.1 and once 
again in Prop.~I.22,  which shows how to construct a triangle with 
sides congruent to given lines.  (The third vertex is the intersection
point of two circles with the specified radii.) Although Euclid
does not explicitly mention the axiom, both its applicability and 
necessity are clear, so we take circle--circle as an axiom.

Line--circle is used only twice in Book I, in Prop.~I.2 and Prop.~I.12, the 
construction of a ``dropped perpendicular.''  We might also 
consider ``one-point line--circle'', in which the conclusion is weakened
to assert only the existence of a single point common to $L$ and $K$.
Since this axiom is inadequate for the application to I.12,  we do 
not consider further the idea of using it instead of line--circle.

Segment-circle has been suggested
as an axiom by many authors, including Tarski (see \cite{tarski-givant}).
But a detailed study shows that it is inadequate; an irremovable
circularity arises in formalizing Euclid without a dimension axiom.
If we try to construct dropped perpendiculars (Euclid~I.12) using 
segment-circle continuity, to check the hypotheses we need
the triangle inequality (I.20).  But  I.19 is needed for I.20, and I.7
for I.19.  In Prop.~I.7, the two triangles that are supposed to coincide might lie
in different planes, but for the hypothesis that they lie on 
the same side of a line, a hypothesis that Euclid stated but 
never used. (He could not have used it, since he never defined
``same side.``)  Even so,
 I.7 is more difficult to prove than Euclid thought, since he took for granted the fact that an angle
cannot be less than itself, but that principle is actually the essential 
content of I.7.  Ever since Hilbert \cite{hilbert1899}, angle inequality 
has been regarded as a defined concept, and proving I.7 then  
requires dropped perpendiculars (I.12) (or at least, we could not 
do without I.12).  But this is circular.
The conclusion is that segment-circle continuity is not a suitable axiom 
to use in formalizing Euclid's proofs.%
\footnote{Line--circle continuity does not suffer
from this problem, as the triangle inequality is not required to drop
perpendiculars.   Of course, as Gupta showed \cite{gupta1965}, one can construct 
dropped perpendiculars without mentioning circles at all, so there is
no formal result that one circle axiom is better for I.7 than another,
as none at all is actually needed.  We merely say that Euclid's proof
can be repaired with line--circle, but not with segment-circle.}

The fact is that each of line--circle and circle--circle implies the other,
in the presence of the other axioms of Euclid. In the interest of 
following Euclid fairly closely, we simply take both as axioms: circle--circle
is used in I.1 and I.22 (triangle construction), while line--circle is used in I.12, and both those proofs
are far simpler than the proofs of line--circle and circle--circle from each other.  

Even though we take both as axioms, we remark on the equivalence proofs.
The proofs can be found in \cite{hartshorne}; see also the last section of 
\cite{beeson-tarski}.  The proof of line--circle from circle--circle relies on 
dropped perpendiculars, which in Euclid is I.12, proved from line--circle.
Therefore, a proof of line--circle from circle--circle must rely instead on 
Gupta's circle-free perpendicular construction \cite{gupta1965, schwabhauser}, 
carrying us far beyond Euclid.  The only known synthetic proof of circle--circle from line--circle uses the ``radical axis'' \cite{strommer1973}.

\subsection{What was Euclid thinking?}
It seems strange that Euclid, who was generally careful, glaringly omits
both line-circle and circle-circle.    
When he needs to use line-circle in the proof of I.2, he instead
says ``Let the straight line $AE$ be produced in a straight line with $DA$''.
In other words, ``let $DA$ be extended until it meets the circle at $E$.''
Remember that lines are always finite, so line-circle intuitively says 
that a line can be extended until it meets the circle,  as well as saying
that (when it is long enough to reach the circle) it cannot pass through the circle at some ``missing point'' without touching.  Probably Euclid 
thought the difficulty was getting the lines long enough,  not getting 
the circle impenetrable. Then he probably had line-circle in mind 
when stating Euclid 2, ``To produce a finite straight line continuously
in a straight line'',  not just ``by some amount'',  and not ``by an
amount equal to a given segment'',  but ``enough to meet a given circle'',
if the starting point is inside that circle.

\section{Equal Figures in Euclid}\label{section:figures}
Euclid defined the word ``figure'' to mean ``that which is 
contained by a boundary or boundaries'',  and explicitly mentioned
that a circle counts as a figure (so boundaries can be curved).  
But in Book I,  figures are triangles and quadrilaterals, so we
do not need to introduce a new primitive sort of variables for ``figure.''
Euclid used the word ``equal'' to denote a relation between figures that 
he does not define.  One possible interpretation is that equal figures are figures with the same area.  But the word ``area'' never occurs in Euclid,
presumably because Euclid realized that he did not know how to define area.
Thus, instead of saying ``the area of the whole is the sum of the areas of the 
parts'', Euclid only reasoned about ``equal figures'', without defining 
that notion.  

Nor did Euclid give any explicit axioms about 	``equal figures''; 
he treated these as special cases of the common notions, such as ``if equals
are added to equals, the results are equal'',  where the ``addition'' of 
figures refers to what we would call the union of disjoint sets.
Occasionally he uses without explicit mention a few further axioms, such as 
``halves of equals are equal.''   

Book~I 
culminates in the Pythagorean theorem,  which Euclid states using the 
notion of equal figures.%
\footnote{In fact, he says that the squares on the sides together equal 
the square on the hypotenuse.  But what he proves is that the square on 
the hypotenuse can be divided into two rectangles, each of which is equal
the square on one of the sides; so the further notion of two figures together 
being equal to a third is not really needed.}
Although we formalized only Book~I in the work reported here,  all the 
propositions in Book~II are about ``equal figures'',  so a correct formulation
of the notion is critical.

Three ways to make Euclid's notion of ``equal figures'' precise have
been proposed in the past.   
First, one can define two figures to be equal if they have equal areas.
 That is not a first-order notion, no matter how area is defined,
  because it involves
measuring areas by numbers.  Second,  after introducing ``segment arithmetic''
geometrically,  one can define area geometrically, but that is a very great
deviation from the path taken by Euclid.

A third interpretation of ``equal figures'' is the
 notion of ``equal content'',  explained on p.~197 of \cite{hartshorne},
which involves cutting figures into a finite number of pieces and reassembling them.
That is also not a first-order notion, because of the ``finite number of pieces'' part.
Hence it is irrelevant for our purposes, and we need not go into the details of the definition.

Conclusion:  the definitions of ``equal figure'' that we find in the literature all suffer 
from one of the following defects:
\smallskip

(i) Not being first order, because of requiring the concept of real number.
\smallskip

(ii)  Not being first order, because of requiring the concept of natural number (even just
for equality of triangle and quadrilaterals).
\smallskip

(iii) First order, but requiring the geometrical definition of coordinates and 
arithmetic (addition and multiplication of finite lines), which goes well beyond Euclid.
\smallskip

The introduction of geometrical arithmetic has already been proof checked in 
\cite{narboux2017}, from Tarski's axioms;  and we checked that Tarski's axioms are 
equivalent to those used in this paper, so approach (iii) has already been proof checked.

We are, of course, not the first ones to face these difficulties.  
Hartshorne lists (p.~196 of \cite{hartshorne}) the properties of ``equal figures'' that
Euclid's proofs use.  Not all the properties in that list are first order.  
  Our approach to the treatment of ``equal figures'' is to 
treat ``equal triangles'' and  ``equal figures'' (that is, equal quadrilaterals) as 
primitive relations, and give first-order
axioms for them, sufficient to account for Euclid's proofs.   These axioms are 
first order versions of Hartshorne's.   

\begin{figure}
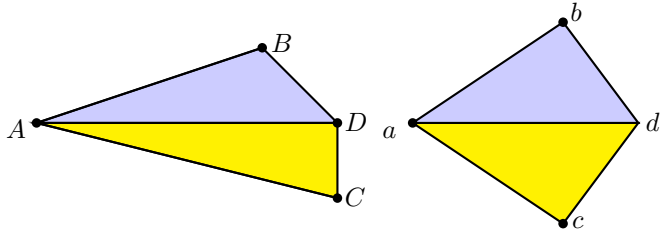

\caption{The axiom {\tt paste3}. If the lower triangles are equal and the upper triangles are equal, then the quadrilaterals are equal.}
\ETPasteFigure
\label{figure:paste3}
\end{figure}

Since we use variables only for points, not for figures, we must use
 two relations: {\tt ET} for ``equal triangles''
and {\tt EF}  for ``equal quadrilaterals'' or ``equal figures''; it is 
only for quadrilaterals, but {\tt EQ} is already taken.
The axioms for these two relations say that {\tt ET} and {\tt EF}  
are equivalence 
relations;  that the order of vertices can be cyclically permuted or reversed,
preserving equality; 
that congruent triangles are equal; that if we cut equal triangles off of 
equal quadrilaterals, producing triangles, the results are equal;  or 
if the cuts produce quadrilaterals, the results are also equal.  Then we 
have ``paste'' axioms that allow for pasting equal triangles onto equal triangles; if the results are quadrilaterals they are equal, provided also 
that the triangles do not overlap, which can be ensured by a hypothesis 
about vertices lying on opposite sides of the paste-line.  Fig.~\ref{figure:paste3} illustrates one of these axioms.  Similarly, if 
pasting equal triangles onto equal quadrilaterals produces quadrilaterals,
they are equal. 

Finally we need an axiom that enables us to prove that certain
figures are {\em not} equal;  all the axioms mentioned so far hold if all figures are equal.
Such an axiom was introduced by de Zolt (see \cite{hartshorne}, p.~201).  But de Zolt's
formulation is not first order.  Instead we take a special case:  if $ABC$ is a triangle,
and $DE$ is a line that cuts the triangle (in the sense that $\B(A,D,B)$ and $\B(B,E,C)$),
then neither of the two resulting pieces $ADE$ or $CDE$ is equal to $ABC$).  This turns 
out to be sufficient.  A complete list of our axioms, including all the equal-figures
axioms, is in the Appendix.

\section{Book Zero and filling in Book I}
We proved more than 230 theorems, including the 48 propositions of Book I. To list 
these theorems in the format used in the Appendix requires 14 pages, and since the files
containing these theorems are accessible (as well as the proofs), we elected not to 
list them all.  Still we wish to give the reader some idea of the additional theorems 
that we had to supply.   We use the phrase ``Book Zero'' informally to encompass those 
theorems that seem to come before Book~I, in the sense of being used in Book~I and not 
depending on Book~I themselves.   Book Zero begins with properties of congruence 
and betweenness; several important and often-used lemmas are about the order of four 
points on a line, when two betweenness relations are known between them.  (There is 
one axiom about that, and the rest of the relevant propositions can be proved.) 
There are variations on the 5-line axiom; there are theorems about collinearity and 
non-collinearity; there is the definition of ``less than'' for finite lines, and 
the ordering properties of that relation and how it respects congruence (or equality)
of finite lines.  Lying on ray $AB$ (which emanates from $A$ and passes through $B$) 
is a defined relation; there are lemmas about how it relates to betweenness and to 
collinearity.  We can ``lay off'' a finite line along a ray, and the result is unique.
Euclid says we can ``add equals to equals'' as a common notion;
the formal version of this is  \verb|sumofparts|, which as we discussed
above can either be an axiom or lemma, depending which version of the 
5-line axiom we take.   There is also \verb|differenceofparts| and
\verb|subtractequals|.   Equality and order of angles are defined concepts, and 
we have to prove their fundamental properties in lemmas such as \verb|ABCequalsCBA|,
\verb|equalanglestransitive|, and so on.  The ``crossbar theorem'' is also part of 
Book Zero.   The notions of ``same side'' and ``opposite side'' are defined, 
and their fundamental properties proved, including the plane separation theorem, according
to which if $C$ and $D$ are on the same side of $AB$, and $D$ and $E$ are on opposite sides
of $AB$, then $C$ and $E$ are on opposite sides of $AB$; that is, there is a point collinear
with $AB$ that is between $C$ and $E$.   This is where we pass out of Book Zero, 
however, since the proof of that theorem requires constructing a midpoint, which 
is Proposition 10. 

 Book Zero comprises about seventy theorems;  Euclid's Book I 
has 48; we proved an additional hundred of so theorems that are needed to prove 
Euclid's 48, or are variants of those propositions,
  but whose proofs use some of Euclid's propositions as well.   Let us 
give a typical example: the lemma we call \verb|collinearbetween| is used 19 times 
in our development, including in Propositions 27,30,32,35,44, and 47.   That lemma
says that if two lines $AB$ and $CD$ are parallel, and there is a point $E$ between 
$A$ and $D$ that is also collinear with $BC$,  then that point $E$ is actually 
between $B$ and $C$.  That is not  trivial to prove, and Euclid simply 
assumes that it is so, because it appears so in the diagram.  
The names of 
some other lemmas will be illustrative: 

\noindent
\verb|droppedperpendicularunique|, 
\verb|angleordertransitive|, 

\noindent
\verb|angleorderrespectscongruence|,
 \verb|angletrichotomy|.   
 
\noindent
Euclid's proof of angle bisection in I.9, via I.1,
cannot be corrected, and instead we prove I.10 (line bisection)
using Gupta's 1965 proof, and then prove I.9.  
Prop.~I.12  (dropped perpendicular) has to 
precede Prop.~I.7 (angle bisection),  because Euclid's proof of 
I.7 is hopelessly inadequate,  and we give a much more complicated
proof that requires perpendiculars.  
  Once perpendiculars are available, we prove
\verb|pointreflectionisometry| and \goodbreak
\noindent\verb|linereflectionisometry|, and use them 
to carry out Szmielew's proof of Euclid's Postulate~4 (all right angles are equal). 
Euclid fails to state \goodbreak
\noindent\verb|legsmallerhypotenuse|, which is needed to prove 
another fact about right triangles that
 Euclid uses without proof:  the foot of the perpendicular from the right angle to 
 the hypotenuse actually lies between the two endpoints of the hypotenuse.  Towards
 the end of Book~I,  the steps of the proofs are more cavalier, and the omitted 
 lemmas are more difficult;  for example Euclid omitted to state and prove that a square
is a parallelogram.%
\footnote{A square has four right angles and equal sides.  A parallelogram has both pairs 
of opposite sides parallel. But there is also this difficulty:  Nothing in 
the definition of a square requires the four sides to lie in the same plane.
That can be (and hence should be) proved.  We discussed the definition
of ``parallel'' in \S\ref{section:definitions}.
}

\section{Formal representation of Euclid}
We wanted to write down our axioms, definitions, postulates, lemmas,
and propositions in a form that would be easy to manipulate by computer,
and independent of any particular computer language, so as to still 
be readable decades or centuries hence.
We chose to use strings to represent all these things. Euclid used 
only one-character variable names, and we did the same.  In that case 
there seemed no need for commas and parentheses; in other words we used
Polish notation.  There were, however, more than 26 relations to consider,
so we used in all cases two-character names for the relations.  For example, 
we write $\B(a,b,c)$ in the form {\tt BEabc}.   $AB = CD$ is written 
{\tt EEABCD}.   We used {\tt EQAB}  to represent $A=B$, so we could not 
use {\tt EQ} a second time, and chose {\tt EE} instead.  ``Parallel'' 
becomes {\tt PR}, as in {\tt PRABCD}.   There are quite a few of these
abbreviated two-character names, but that is enough to convey the idea.
The point is that every formula is a string.  Conjunctions and disjunctions
begin with {\tt AN} or {\tt OR} and the subformulas are separated by 
{\tt +} or {\tt |}, respectively.  Negations are formed with {\tt NO}.

Then we define classes {\tt Axiom}, {\tt Definition}, and {\tt Theorem},
each of which has fields called {\tt label}, 
{\tt hypothesis}, {\tt conclusion}, and {\tt existential}.
The {\tt label} field is used for the name.  The {\tt hypothesis} 
and {\tt conclusion} fields each contain an array of formulas, or 
a single formula.   The {\tt existential} field contains  an empty string by 
default, and if it isn't empty, that means that it contains a list of 
variables that are supposed to be existentially quantified in the conclusion.

We have thus defined a subset of first-order predicate logic.  Specifically,
our formulas have no function symbols,  and only existential quantifiers; 
universal quantification over the free variables is left implicit.  Nested
quantifiers do not occur.   Polish notation, one-character variables, and 
two-character predicate names make it easy to manipulate these formulas as strings
and arrays of strings,  and substitutions can be coded as arrays, making 
unification possible by regular-expression matching.  Every modern programming 
language has useful libraries for this sort of thing.

\section{Formal proofs}
Each proof is a list of lines.  Each line contains a formula and 
optionally a justification.  The proof is kept in a {\tt .prf} file
whose name gives the label of the theorem it is intended to prove. 
The proof begins with a sequence of unjustified lines that must 
repeat the hypotheses of the theorem.   It ends with a line that is 
the conjunction of the conclusions of the theorem,  or the sole 
conclusion if there is only one.  The first line after the hypotheses 
must have a justification.  Any unjustified lines in the rest of the 
proof must either be repetitions of earlier lines or must follow 
by logic alone from some earlier lines.  Justifications follow the 
pattern {\tt kind:label}, where {\tt kind} is either {\tt defn},
{\tt axiom}, {\tt postulate}, {\tt proposition},
 or {\tt lemma}, and {\tt label} is 
the label of an item of the specified kind.   An axiom, postulate, or 
definition can be used anywhere, but a lemma or proposition cannot 
be used anywhere,  because circular arguments must be prevented.

We avoid circularity by having a ``master list'' of lemmas and propositions,
which are 
to be proved in the specified order.  A valid proof (of a certain item in the 
master list) is only allowed to reference previous items in its 
justified steps.

Euclid's proofs,  and ours,  make use of arguments by contradiction 
and cases.  We will now explain the syntax we used.   An argument 
by contradiction is introduced by a line with the justification 
{\tt assumption}.   After some steps of proof, this line must 
be matched by a line with the justification {\tt reductio}. This 
line must contradict the assumption line.   By saying that $A$ and 
$B$ contradict each other, we mean that one of them is the negation 
of the other.  We found it helpful to indent the lines between the 
assumption and the reductio labels, especially when nested arguments 
by cases or reductio occur.

The syntax for cases and proof by contradiction is as illustrated in Fig.~\ref{figure:prooffragment}.  (The example chosen is the
proof of Playfair's form of the parallel postulate, discussed in 
\S\ref{section:euclid5}.)  The first two lines state the hypothesis, that 
we have two lines $CD$ and $CE$ through $C$, both parallel to $AB$.
The last line states the desired conclusion, that $C$, $D$, and $E$
are collinear, i.e., the two parallel lines coincide. In the proof, 
{\tt BE}  means ``between'' and {\tt CR}  means ``crosses'', in the 
sense given in the line labeled {\tt defn:cross}. 

We wrote more than two hundred thirty formal proofs in this syntax.   For convenience,
we introduce a name for this subset of first-order logic:  {\bf Euc},  the first three
letters of Euclid.   We believe that these proofs will be readable,
and computer-checkable,  into the indefinite future; in particular, past 
the lifetime of the currently popular proof assistants that we used to 
check the correctness of these proofs. 

Polish notation is not the favorite of some people, including our referee.
But its durability over a long future seems assured--after all it is already
more than a century old.  We are sure that the scientists of the future will 
easily be able to add parentheses and commas if they prefer to look at 
the formulas with those additions;  of course they could equally easily 
strip out the commas and parentheses, so we think it is not an important 
issue.  With or without parentheses and commas, we believe a low-level
simple formalism is preferable to a high-level language (for example the 
Mizar language), because the purpose of these proofs is to 
permit easy automatic translation into any desired formal language, without
the necessity of resurrecting a language used in the distant past.  It is
not our purpose in this paper to make the proofs easily readable by humans.  That subject will be taken up separately; but our intention is again that 
formal Polish-notation proofs be the starting point of a translation into
something that humans like to read.  
 
\begin{figure}
\caption{An illustrative proof, showing the syntax of formulas and proofs.}
\label{figure:prooffragment}
\begin{verbatim}
PRABCD
PRABCE
NOORCRADBC|CRACBD  assumption 
 ANNOCRADBC+NOCRACBD
 NOCRADBC
 NOCRACBD
 PRABDC  lemma:parallelflip
 CRACDB lemma:crisscross
 ANBEApC+BEDpB  defn:cross
 BEApC
 BEDpB
 BEBpD  axiom:betweennesssymmetry
 ANBEApC+BEBpD
 CRACBD defn:cross
ORCRADBC|CRACBD reductio
cases COCDE:CRADBC|CRACBD
 case 1:CRADBC
  COCDE lemma:Playfairhelper2
 qedcase
 case 2:CRACBD
  ANBEApC+BEBpD defn:cross
  BEApC
  BEBpD
  ANBEBpD+BEApC
  CRBDAC defn:cross
  PRBACD  lemma:parallelflip
  PRBACE   lemma:parallelflip
  COCDE lemma:Playfairhelper2
 qedcase
COCDE cases
\end{verbatim}
\end{figure}

\section{Checking the proofs by computer}

The proofs described in this paper only need a rather weak logic.
There are no function symbols, and all the statements have
a very restricted form.
That made it easy to write a custom-built proof checker, 
or ``proof debugger'',  that we used while developing
the formalization.  That tool also
checks that we stay within the bounds of that logic.

HOL and Coq
use an architecture that guarantees a much higher
reliability.  

This is called the \emph{LCF architecture}, after the LCF system
from the seventies that pioneered the approach.
This architecture divides the system in a small kernel
(or \emph{logical core}) and the rest of the code.
By the use of abstract datatypes, the correctness of the mathematics
is then guaranteed by the correctness of the kernel.
Whatever errors the rest of the code of
the system may contain, the statements claimed to have been proved
will indeed have been proved.
In the case of the HOL Light system (one
of the incarnations of HOL), the correctness of this kernel
has even been \emph{formally} proved (using the HOL4 system, another incarnation of HOL),
which gives an \emph{extremely} high guarantee
that the system will not have any logical errors.

Our procedure for proof-checking Euclid was thus as follows:

\begin{itemize}
\item  Write formal proofs in the \Euc\ language,  simultaneously 
checking and debugging them with our custom proof debugger.

\item  Translate these proofs into HOL Light or Coq syntax
by means of simple scripts.

\item  Check the resulting proofs in HOL Light and Coq
\end{itemize}

Was it necessary to use the \Euc\ language?  No,  but it is closer
to Euclid than either HOL Light or Coq,  and allowed us to write the 
proofs only once,  and moreover, has a better chance of being readable
a thousand years from now.  Was it necessary to write a custom
checker or debugger for \Euc?   Perhaps not, but it facilitated our 
work flow by separating the writing of the proofs from the use of the
two major proof assistants, and by providing very useful error messages
in the case of incorrect proofs.  Was it necessary to use {\em two}
proof assistants?  No, since each one is perfectly reliable, but we 
did it anyway.  

The devil might ask whether we have lost something 
by using higher order logic to check first order proofs.  
While it may appear so at first glance, actually higher order logic 
itself ensures that we have not.  Consider:  in higher order 
logic (both in Coq and HOL Light) we proved that for any type of points, and 
any predicates satisfying the axioms of \Euc{}, all our theorems are 
satisfied.   In these statements
there is a second-order quantification over the predicates used to 
interpret betweenness, congruence, and the predicates $ET$ and $EF$.
  In essence we have proved 
that the axioms hold in any model (of our axioms for geometry).
Then by G\"odel's completeness 
theorem,  the theorems are actually first order theorems.
However, it is a general feature of higher order theorem provers
that they do not directly check first-order proofs.  Moreover, we 
used the Leibniz definition of equality,  so indeed our translated
proofs are {\em not} first order.  However, we did check the first order
proofs directly in our custom proof checker before translating them 
to higher order logic.    

 Our debugger also counted the number of inferences.  
  Proofs of more than 200 inferences were not uncommon,
but the majority were under 100 inferences.

We also wrote code that analyzed the dependencies between lines of a given 
proof.  This enabled us to identify and eliminate lines that were never
subsequently used.  We follow Euclid in sometimes repeating previously deduced 
lines just before applying a proposition, to make it apparent that the 
required instances of the hypotheses have indeed been derived.  These lines,
of course, are technically eliminable; but mainly we wanted to eliminate 
``red herring'' lines that were actually irrelevant.  The automatic
detection of such lines was useful.

\section{Checking the proofs in HOL Light}

To ensure the correctness of the Euclid formalization from the language
\Euc\ to HOL Light, we built a
very small custom checker on top of HOL Light, and used that to check our work
for correctness in HOL as well.
The source of this proof checker,
\texttt{proofs.ml},
has about a hundred lines,
which then is used to check a translation of the formalization into syntax that HOL
Light can process of almost twenty thousand lines, \texttt{michael.ml}.
This last file is created from the original proof files described
above by two small \emph{ad hoc} scripts, a PHP script called \texttt{FreekFiles.php}
and a Perl script called \texttt{FreekFiles.pl}.

In the HOL system, all input (even the proofs) always consists of executable ML source code.
The proofs of each lemma in our case is checked by calling a function \verb|run_proof|
(implemented in \texttt{proofs.ml}) on the statement of the lemma
and a list of items of a custom datatype called \texttt{proofstep}.
The ML definition of this datatype is shown in Fig.~\ref{figure:proofstep}.
\begin{figure}
\caption{The ML datatype used when checking the proofs with HOL.}
\label{figure:proofstep}
\begin{verbatim}
type proofstep =
  | Known of term
  | Step of term * thm
  | Equalitysub of term
  | Assumption of term
  | Reductio of term
  | Cases of term * term
  | Case of int * term
  | Cases_ of term
  | Qedcase;;
\end{verbatim}
\end{figure}
The output of this function is a HOL \texttt{thm}, a proved statement.
In other words, the ML type of the function that is used here is:
\begin{center}
\verb|run_proof : term -> proofstep list -> thm|.
\end{center}
The list of \texttt{proofstep}s corresponding to the proof 
from Fig.~\ref{figure:prooffragment} is shown in Fig.~\ref{figure:prooffragmentHOL}.
\begin{figure}
\caption{The HOL counterpart of the proof 
in Fig.~\ref{figure:prooffragment}.}
\label{figure:prooffragmentHOL}
\begin{verbatim}
let lemma_Playfair = run_proof
  `PR A B C D /\ PR A B C E ==> CO C D E`
  [
    Known `PR A B C D`;
    Known `PR A B C E`;
    Assumption `~(CR A D B C \/ CR A C B D)`;
    Known `~CR A D B C /\ ~CR A C B D`;
    Known `~(CR A D B C)`;
    Known `~(CR A C B D)`;
    Step (`PR A B D C`, lemma_parallelflip);
    Step (`CR A C D B`, lemma_crisscross);
    Step (`BE A p C /\ BE D p B`, defn_cross);
    Known `BE A p C`;
    Known `BE D p B`;
    Step (`BE B p D`, axiom_betweennesssymmetry);
    Known `BE A p C /\ BE B p D`;
    Step (`CR A C B D`, defn_cross);
    Reductio `CR A D B C \/ CR A C B D`;
    Cases (`CO C D E`, `CR A D B C \/ CR A C B D`);
    Case (1, `CR A D B C`);
    Step (`CO C D E`, lemma_Playfairhelper2);
    Qedcase;
    Case (2, `CR A C B D`);
    Step (`BE A p C /\ BE B p D`, defn_cross);
    Known `BE A p C`;
    Known `BE B p D`;
    Known `BE B p D /\ BE A p C`;
    Step (`CR B D A C`, defn_cross);
    Step (`PR B A C D`, lemma_parallelflip);
    Step (`PR B A C E`, lemma_parallelflip);
    Step (`CO C D E`, lemma_Playfairhelper2);
    Qedcase;
    Cases_ `CO C D E`;
  ];;
\end{verbatim}
\end{figure}

In that example some of the choices on how to
translate the statements from the proof to the HOL logic can be seen.
For instance, we had to decide whether to translate \texttt{EQ}
to the standard HOL equality, or to have it be a custom relation.
We chose to make use of the features of standard first order logic
with equality, but nothing beyond that.
That means that we translated equality to the built-in equality of the
logic, and translated \texttt{NE} identical to \texttt{NOEQ}
and \texttt{NC} identical to \texttt{NOCO}.
Therefore
the translation does not have predicates \texttt{EQ}, \texttt{NE} and \texttt{NC}.
As a small optimization, \texttt{NONC} was translated without a
double negation.

Within HOL we used an axiomatic approach.  That is,
we added our axioms to the HOL axioms.  That way, we 
will be verifying that the theorems of Euclid follow from 
those axioms,  rather than (for example) that they are true
in $\R^2$ or $\R^n$.
This raised the number of HOL axioms by 36, from the original 3 to 39.
Before stating these 36 axioms we also added two new primitive types. \texttt{point}
and \texttt{circle}, and five new primitive predicates, \texttt{BE}, \texttt{EE},
\texttt{CI}, \texttt{ET} and \texttt{EF}.

There are some differences between what was taken to be axioms in the
original version of the formalization, and what are axioms in the
HOL version.
The definitions of the predicates were originally axioms, but in the
HOL version are actual HOL definitions, with the `axioms' being the equivalences
that these definitions produce.
Two exceptions for this approach are \texttt{defn:unequal} and \texttt{defn:circle}
of which the first is omitted (it is not used anywhere) and the
second is an axiom (as it does not have the shape of a HOL definition).
Also, \texttt{defn:inside}, \texttt{defn:outside} and \texttt{defn:on}
are still axioms, because these also state that the
defining property does not depend on the points that give the circle,
which one does not get from just a definition.

Of the common notions, \texttt{cn:equalitytransitive},\goodbreak
\noindent\texttt{cn:equalityreflexive}, \texttt{cn:stability} and
\texttt{cn:equalitysub} are not axioms, but proved statements,
because they only involve equality and are part of the logic.

The LCF architecture of HOL Light only guarantees that the proofs
are valid in the higher order logic of HOL Light.
However, the implementation in \texttt{proofs.ml} only uses first
order tactics (most notably \texttt{MESON}, which is the main
tool used for checking the steps), which shows that the proofs
are actually first order.
The proofs, written in the {\Euc} language, contain on each line
a formula and its justification.  HOL Light checks that the 
formula follows from previous steps of the proof, using the 
axiom, theorem, or definition mentioned in the justification, 
together with first-order logic.  The simple logic associated 
with {\Euc} allows only instantiation of the axiom, theorem, or 
definition; or in proofs by cases and contradiction, some simple
propositional logic.  Is HOL Light then really checking the 
given proof?  Yes, it is, because (a) even if it found a 
more complicated first-order proof, that would still be a check, 
and (b) in any case, HOL Light will not use quantificational logic 
to prove a theorem without quantifiers.

The full check of the twenty thousand lines is uneventful, but
not very fast, mostly because of the use of the rather heavy \texttt{MESON}.
It takes several minutes.
At some points \texttt{MESON} has to work very hard (it loses the fact
that the conclusion of the lemma exactly matches the step being proved, and is
trying many possible ways to unify the parts).
For these cases \texttt{proofs.ml} contains two custom lower level tactics,
\verb|SUBGOAL_UNFOLD_TAC| and \verb|SUBGOAL_MATCH_TAC|,
which do not use \texttt{MESON}.

\section{Checking the proofs in Coq}

\begin{figure}
\begin{verbatim}
Lemma lemma_Playfair : 
   forall A B C D E, 
   Par A B C D -> Par A B C E ->
   Col C D E.
Proof.
intros.
assert (neq A B) by (conclude_def Par ).
assert (neq C D) by (conclude_def Par ).
assert (~ ~ (CR A D B C \/ CR A C B D)).
 {
 intro.
 assert (Par A B D C) by (forward_using lemma_parallelflip).
 assert (CR A C D B) by (conclude lemma_crisscross).
 let Tf:=fresh in
 assert (Tf:exists p, (BetS A p C /\ BetS D p B)) by (conclude_def CR );
    destruct Tf as [p];spliter.
 assert (neq D B) by (forward_using lemma_betweennotequal).
 assert (neq B D) by (conclude lemma_inequalitysymmetric).
 assert (BetS B p D) by (conclude axiom_betweennesssymmetry).
 assert (CR A C B D) by (conclude_def CR ).
 contradict.
 }
assert (Col C D E).
by cases on (CR A D B C \/ CR A C B D).
{
 assert (Col C D E) by (conclude lemma_Playfairhelper2).
 close.
 }
{
 let Tf:=fresh in
 assert (Tf:exists p, (BetS A p C /\ BetS B p D)) by (conclude_def CR );
    destruct Tf as [p];spliter.
 assert (CR B D A C) by (conclude_def CR ).
 assert (Par B A C D) by (forward_using lemma_parallelflip).
 assert (Par B A C E) by (forward_using lemma_parallelflip).
 assert (Col C D E) by (conclude lemma_Playfairhelper2).
 close.
 }
(** cases *)
close.
Qed.
\end{verbatim}
\caption{Example proof in Coq}
\label{examplecoq}
\end{figure}

\subsection{Formalizing the Axioms}

The axioms in our axiomatization of Euclid are of two kinds: axioms related to definitions and others. Axioms that serve as definition as those of the form:
 $\forall x, P(x) \iff Q(x).$  
 We translate them to a proper Coq definition to reduce the number of axioms.
 Technically, in Coq, we did not use the Axiom keyword. Because axioms are similar to global variables in a programming language, 
 they reduce the re-usability of the code. 
We sorted the axioms into several groups, and defined them in Coq using type classes. Then, the axioms are given as so-called section variables of Coq, a mechanism which allows to have the axioms as an implicit assumption for each lemma. 
For propositions I.1 to I.28 and I.30 we do not use the fifth postulate of Euclid. For propositions I.1 to I.34, we do not need the equal-figure axioms. 
Avoiding the Axiom keyword  
allows us to reuse the proofs in a different setting by proving the axioms in Coq as a second-order property either from another axiom system or by constructing a model (see~Sec.\ref{verifying-axioms-coq}).

\paragraph*{Equality}

We model the equality using Coq's built-in equality: Leibniz's equality.
We could also have assumed an equivalence relation and substitution properties for each of the predicates of the language.

\subsection{Verifying the proofs}

Proof assistants differ in their mathematical foundations (e.g.\ type theory, higher order logic (HOL), or set theory) and their proof language. In procedural style proof assistants (e.g.\ Coq and HOL Light), proofs are described as a sequences of commands that modify the proof state,  whereas in proof assistants that use a declarative language (e.g\ Mizar and Isabelle), the proofs are structured and contain the intermediate assertions that were given by the user and justified by the system. 

We wrote a script to translate the proofs to Coq's language. The translation is relatively easy as the proofs steps used by the proof debugger are small.

Our translation generate a proof in the traditional language of Coq, not in the declarative language introduced by Corbineau~\cite{corbineau2008declarative} because this language is not maintained. But, the formal proofs we generate are in the declarative style: for case distinctions we give explicitly the statement which is used instead of the name of the hypothesis, the proofs are purely in the forward chaining style, based on sequences of applications  of the standard Coq tactic \texttt{assert}.
We do not introduce hypothesis numbers. 
The existential statements and conjunctions are eliminated as soon as they appear by introducing the witness and decomposing the conjunction.
The assertions are justified using some ad-hoc tactics which use some automation and congruence closure.
We had to circumvent some weakness of Coq's automation. Coq is not able to use efficiently lemmas of the form $ \forall x y z, P x y z \rightarrow P y z x \land P y x z $, because the \texttt{apply} tactic will always choose to unify the goal with the first term of the conjunction. The standard way to state such a lemma in Coq is to split the lemmas in two parts. But, we did not want to modify the original formalization of the lemmas. Hence, to circumvent this limitation, we verified these proof steps using a tactic based on forward-chaining. 
The case distinction tactic allows to distinguish cases on previously proved disjunctions or disjunctions which are classical tautologies. The tactic can deal with n-ary disjunctions. The proof script is structured using curly brackets and indentation.
Each proof step of the original proof corresponds to one proof step in Coq, except that steps for which there is no justification in the original proofs. Those steps correspond to the natural deduction rule for introduction of implication and are implemented using the standard Coq tactic \texttt{intro}.
All tactics are designed such that the names of the geometric objects are preserved from the original proofs but the names of the hypotheses are automatically generated by Coq and not used explicitly in the proofs. 
Figure~\ref{examplecoq} displays the proof of Playfair's axiom of uniqueness of parallels in Coq's language enriched by the tactics to ease the verification. Some of the predicates have been renamed into longer names to enhance readability and to match the names used in the GeoCoq library.
The proofs are verified using classical logic.
The proofs can be checked by Coq in about 90 seconds using an
Intel(R) Core(TM) i7-7700 CPU @ 3.60GH with 32Go RAM.

\subsection{Verifying the statements}

When evaluating a formalization, even if we trust the proof checker, we need to check that statements are formalized faithfully. Usually the only method we can use for this process is to check the statements by human inspection and trust the reviewer to also check the statements. For this formalization, we were lucky, as many statements had been formalized independently by the first author and the GeoCoq team. To improve the confidence in the formalization, we compared the two formalizations of the statements to detect potential defects.
We detected only minor differences in some of the statements.

\section{The axioms hold in $\R^2$}
The axioms fall into two groups:  those that are variants of 
Tarski's axioms A1--A10,  and the equal-figure axioms.
In order to make sure that there is no mistake in the axiomatization,
we wished to check formally that the axioms hold in the Cartesian
plane $\R^2$.   For reasons of convenience, we checked that for the first 
group of axioms in Coq, and for the equal-figure axioms in HOL Light.
We will discuss these two verifications separately.

It has already been checked \cite{narboux2017b} that Tarski's axioms
hold in $\R^2$.   In Tarski's A1--A10, without dimension axioms,
 one can (formally) verify
our axioms, which are mainly different from Tarksi's by using strict 
betweenness and hence avoiding degenerate cases.  This has also been 
verified in Coq. 

\subsection{Verifying the Tarski-style axioms in Coq}
\label{verifying-axioms-coq}
Since there is no 
dimension axiom,  the ``intended model'' is $\R^n$,
for any integer $n > 1$.  More generally, 
we wish to prove that if $\F$ is a Euclidean field,
then $\F^n$ satisfies our axioms.  We break this 
claim into four parts.   

(i) our axioms for
neutral geometry can be derived from the corresponding Tarski's axioms.

(ii) circle-circle, circle-line, and Euclid 5 can be derived from the corresponding axioms from Tarski.

(iii) Tarski's axioms hold in $\F^n$.  By ``Tarski's axioms`` we mean
those of our axioms that expressed using betweenness and equidistance.
(These are similar to the axioms and some theorems of Tarski's geometry,
except we use strict betweenness.)

(iv) the equal-figure axioms hold in $\F^n$.

The reason for passing through Tarski's axioms
is that it has already been shown that 
Tarski's axioms for neutral geometry hold
in $\F^2$, when $\F$ is a Pythagorean ordered
field.
\footnote{Recall that a Pythagorean field, is a field where sums of squares are squares.} 
 Specifically, 
  Boutry and Cohen have formalized the proof that the Cartesian plane over a Pythagorean ordered field is a model of our formalization of Tarski's axioms 
 A1--A10 using ssreflect~\cite{boutryphd}.

Ad (ii).  We chose to assume the circle-circle intersection property  and use the formal proof obtained by Gries and the second author that the circle-line intersection property can be derived from circle-circle intersection, even without assuming a parallel postulate~\cite{griesinternship}.
Gries also formalized the proof that circle-circle intersection can be derived from the continuity axiom of Tarski (Dedekind cuts restricted to first-order definable sets).
As \Euc{ } axiom system assumes that we have a sort for circles, we need to define the type of circles from Tarski's axioms (and Coq's Calculus of Inductive Constructions CIC). The type of circles is defined as the triple of points (\texttt{*} is interpreted in this context as the Cartesian product).
\begin{verbatim}
Definition Tcircle : Type :=  Tpoint*Tpoint*Tpoint %type.
\end{verbatim}
Then the predicate {\tt CI} can be defined by:
\begin{verbatim}
Definition CI (J:Tcircle) A C D := J=(A,C,D) /\ C<>D.
\end{verbatim}
Then the relation expressing that a point is on a circle can be defined by destructing the triple:
\begin{verbatim}
Definition OnCirc P (C:Tcircle) := 
  match C with 
  (X,A,B) => tarski_axioms.Cong X P A B
  end.
\end{verbatim}

For Euclid 5, we rely on the proofs of equivalence between different versions of the parallel postulates studied previously by the first author~\cite{beeson-bsl} and formalized  in Coq, by Boutry, Gries, Schreck and the second author~\cite{narboux2017c}.
 
\subsection{Verifying the equal-figure axioms}

We here outline the verification that the equal-figure axioms
hold in $\R^2$.  We did not verify them in $\F^n$ or even $\R^n$,
because we wanted to use the (scalar) cross product in $\R^2$ and 
the existing tools for real and vector algebra in HOL Light.

We interpret points as members of the HOL Light type \verb!real^2!.
We write that here as $R^2$.  Then we define the dot product 
and the two-dimensional cross product as usual:
\begin{eqnarray*}
(a,b) \cdot (c,d) &=& ac + bd \\
(a,b) \times (c,d) &=& ad-bc \\
\end{eqnarray*}
Twice the signed area of a triangle $abc$ is defined by 
$$ tarea(a,b,c) = (c-a) \times (b-a). $$
We define $ET$  (equal triangles) by saying that 
two triangles are equal if the absolute values of their signed areas are equal.
That is, $ET(a,b,c,p,q,r)$ means
$$\vert tarea(a,b,c) \vert = \vert tarea(p,q,r) \vert.$$
Twice the signed area of a quadrilateral $abcd$ is given by the cross 
product of its diagonals:
$$ sarea4(a,b,c,d) = (c-a) \times (b-d). $$
The area of $abcd$ is the absolute value of the signed area.
We wish to define $EF$ (equal quadrilaterals) by saying two 
quadrilaterals are equal if the squares of their signed areas are equal.
But we do not allow just any four points to be a quadrilateral. Instead
we allow two kinds of quadrilateral, convex quadrilaterals and ones 
which are ``really triangles``,  meaning that one vertex is between two 
adjacent vertices.  We define $abcd$ to be convex if its diagonals cross,
i.e., there is a point $m$ with $\B(a,m,c) \land \B(b,m,d)$. In our 
formal development the condition $0 < area(a,b,c,d)$ is imposed separately.
We interpret $EF(a,b,c,d,p,q,r,s)$ to mean that $abcd$ and $pqrs$ are
each quadrilaterals in this sense, and their areas are equal. 

Betweenness $B(a,b,c)$ is  
interpreted (or defined) thus:
\begin{eqnarray*}
B(a,b,c) &\leftrightarrow& \exists t. (b-a) =(c-a) > 0 \land  0 < t < 1
\end{eqnarray*}

These are all the definitions needed to interpret the equal-figures axioms
in $R^2$.  We executed this translation by hand,  producing a list
of sixteen goals to prove in HOL Light.  These theorems turned out 
not to be as trivial as we initially thought;  the axiom that says 
congruent triangles are equal required nearly 2000 lines of HOL Light proof.
Even more lines were required for the last axiom, {\tt paste4}.  Altogether
the verification of all 16 equal-figure axioms 
required about six thousand lines of HOL Light proofs.

Although theoretically these theorems fall within the domain of 
quantifier elimination for the real field, in practice they have too 
many variables, and quantifier elimination was not used.  Instead,
we used rotations and translations to reduce
the complexity, and used only the standard vector-algebra theorems
that are distributed with HOL Light.    Many of the axioms are
consequences of the additivity of area, so we proved them by 
first proving (four or five different forms of) the additivity of area, and then deriving the 
axioms from the additivity of area. Numerous lemmas seemed to be required.
Just to mention two examples:  

(i) {\tt between\_norm}, which says that $B(a,b,c)$ is 
equivalent to the ``norm condition'', 
$$\vert b-a \vert + \vert c-b \vert = \vert c-a \vert.$$

(ii) If $abcd$ is a convex quadrilateral (its diagonals meet), and has positive area, then $ab$ and $cd$ have no point in common.

As mentioned, we used the scalar cross-product to define area.  Alternately,
we could have used the definition of area by Lebesgue measure, which already
exists in the HOL Light library.  Since we did not do that, we cannot say
for certain that it would not have been easier, but there certainly 
would have been many details and extra lemmas in that approach as well.

\subsection{An inconsistency and its repair}

Our equal-figure axioms include axioms about cutting and pasting 
equal figures to get other equal figures.  See Fig.~\ref{figure:paste3}
for an example, namely {\tt paste3}, which 
is about pasting together triangles with a common 
side $AC$ to get a quadrilateral $ABCD$.  This is used in 
Propositions 35, 42, and 47.  Of course we need something to ensure that 
the two triangles are in the same plane, and that when pasted together,
they do make a quadrilateral. The obvious hypothesis is that 
$B$ and $D$ are on opposite sides of the common side $AC$, and that was our 
first formulation. 
 
Our equal-figure axioms also include axiom {\tt cutoff2}, which
says that if we cut equal triangles off of equal quadrilaterals, with one 
end of the cut at a vertex and the other on a non-adjacent side, the results
are equal quadrilaterals. This axiom 
is used only once, in Prop.~43; that step of Euclid's proof is justified 
by common notion 3,  about subtracting equals from equals. 
We formulated this axiom without any formal statement that the resulting
quadrilateral is in fact a quadrilateral;  nor does Euclid have any 
such justification in his proof.  

It turned out, as we discovered when attempting to verify that the axioms 
hold in $\R^2$, that these two formulations are inconsistent:  This 
version of {\tt paste3} permits the creation of a non-convex quadrilateral
$ABCD$ in which diagonal $BD$ passes outside the quadrilateral.  Then 
when we attempt to cut off a triangle $BAP$ with $P$ on $AD$, the triangle
is outside rather than inside, so the area of the result might be larger,
rather than smaller, than the area of $ABCD$.  This phenomenon leads to 
an inconsistency. 

There are two possible ways to remedy this problem:  either strengthen 
the hypotheses of {\tt cutoff2} (requiring that the resulting quadrilateral
have crossing diagonals),  or strengthen the hypotheses of {\tt paste3}
(requiring that $ABCD$ have diagonals that cross, or meet at $A$ or $C$).
Either way works.  Which is more faithful to Euclid?  Since these axioms 
are used only a few times in Euclid,  it is not a matter of much importance.
The argument against modifying {\tt cutoff2} is that we would then have
to verify the added condition in the proof of Prop.~43,  which Euclid
evidently felt no need to do, and which would probably double the length
of the proof of Prop.~43. The argument against modifying {\tt paste3} is that
it might not suffice for a proof that requires constructing a non-convex
quadrilateral.  But there are no non-convex quadrilaterals in Books I-III,
so we chose to modify {\tt paste3} rather than {\tt cutoff2}.  This
permits us to apply these axioms in Euclid's proofs without adding 
more steps.

\section{Previous work on computer checking geometry}\label{section:history}

Work on computerizing Euclidean geometry began in 1959, 
in the first decade of the computer age, 
with the pioneering work of Gelernter \cite{gelernter1959, gelernter1960}.
(The reference has a later date because it is a reprinting in a collection.)
The axiom system used by Gelernter was not given explicitly,
but from the example proofs given, it can be seen that it 
was, at least in effect, a points-only system.  It was a strong 
axiom system, including for example all the triangle congruence 
theorems, Euclid 4,  some strong but unspecified betweenness axioms.
Tarski is not referenced; that is not surprising as Tarski's first publication
of his axiom systems was also in 1959. Gelernter's system was 
claimed to be as good as ``all but the best'' high school geometry 
students.  Considering the primitive hardware and software of 1959,
it was an amazing program.   However, its authors stated that 
they viewed geometry as just one area in which to study heuristic
reasoning, and neither the program nor its underlying formal theory
ever raised its head above water again.

As far as we know, nineteen years passed before the next work 
in computerizing geometry; it was 1978 when Wen-Ts\"un Wu \cite{wu1978} began 
a series of papers on the subject, culminating in his 1984 book, 
published only in Chinese, and not available in English until 1994 \cite{wu1994}.
 He was soon assisted by S.~C. Chou \cite{chou1987,chou1988}. Using  
 coordinates, one reduces
 a geometry theorem to an implication between polynomial equations.
One can then demonstrate the truth of a geometry theorem by algebraic 
methods.  Such algebraic methods, 
while they may succeed in establishing the validity of a theorem,  do 
not provide a proof from geometric axioms.  One might check the 
correctness of Euclid's {\em results} this way, but not his {\em proofs.}
One problem with this approach is that it works only for problems that involve
equality, not for problems that involve betweenness or inequality.  
A second problem is that
Wu's method requires polynomials with hundreds of variables, and hence does not produce
human-comprehensible proofs.   Nevertheless, Wu and Wang, Gao,
Chou, Ko, and Hussain proved many theorems.  Within a few years three different 
groups began to use Gr\"obner bases to do the algebraic work, instead of the Wu-Ritt algorithm. 
See \cite{kapur1988}, \cite{kutzler1986}, \cite{chou-schelter}.

Sometime after 1984, Chou invented the {\em area method},  which still 
uses polynomial computations, but based on certain geometric invariants.  The 
area method, as an algorithm for solving geometry problems, can be used by humans 
and has even been used to train students for Olympiad-style competitions. \cite{chou1994}, p.~xi.
For the state of the art in area-method implementation as of 1994, see \cite{chou1994},
where more than four hundred computer-produced and human-readable proofs are given.
However, the steps of these proofs are equations, whose truth is verified by 
symbolic computation, not by logic. Also, as with Wu's method, inequalities and 
betweenness cannot be treated.   

In the 1980s, Larry Wos experimented with proving geometry theorems 
in Tarski's theory using the theorem-prover {\textsc OTTER}.  Then Art Quaife
took up that same project, publishing a paper in 1989 \cite{quaife1989},
and devoting a chapter of his 1992 book to it.  Wos and Quaife left a 
number of ``challenge problems'' unsolved by {\textsc OTTER}.   In this 
same decade, the book \cite{schwabhauser} was published, containing the 
results from Gupta's thesis \cite{gupta1965} and Szmielew's Berkeley course,
together constituting a systematic development of ``absolute geometry'' 
(no circle or continuity axioms) from Tarski's axioms.  This book 
was quite rigorous, but not (yet) computer-checked. It also did not 
reach even to the beginning of Euclid.  Twenty years later, the first author
 and Wos returned to this project, and used {\textsc OTTER} to find proofs of 
all the challenge problems of Quaife, and the first ten chapters of 
\cite{schwabhauser} (Part I).  But since \cite{schwabhauser} spends
a lot of effort developing ``elementary'' results from minimal axioms
(no circle axioms and postponing the parallel postulate as long as possible),
the propositions of Euclid are not reached.  In this project, the more difficult
proofs were not found automatically, but instead the theorem prover was
used almost like a proof checker, by means of supplying ``hints.''  
Therefore, when we wanted to proceed to proof check Euclid, it 
seemed appropriate to switch from a theorem prover to a proof checker,
which is what we did for this work.

In 2009,  Avigad, Dean, and Mumma \cite{avigad2009} reported on a formal system 
for Euclid's {\em Elements}.  This system is six-sorted (points, lines, 
circles, segments, angles, and ``areas'' (figures), and therefore has also
a large number of primitive relations, including ``same side'' and ``opposite side''.
The axiom system of E differs from ours, because it contains more axioms. They assume 20 constructions axioms, 34 axioms about the two-side, inside and betweenness relations which they call ``diagrammatic inferences.''  Their system is intended for two-dimensional geometry only, a
restriction deliberately avoided in our system.  
They also assume that distances and areas can be {\em measured} using a linearly ordered abelian group. We follow Euclid in not assigning measures to distance or area;  one 
may compare distances or areas (figures), but not measure them.

 The intention 
of those authors (see \S6 of \cite{avigad2009}) was to build an interactive proof checker (with application to 
education).  They wanted to separate the ``diagrammatic inferences''
from the inferences that Euclid wrote out, using an algebraic program 
for the diagrammatic inferences and a prover for the others. The steps
that Euclid omitted would be done by algebra instead of by logic.
This is a ``hybrid'' approach, halfway between the completely axiomatic
approach and the computer-algebra techniques of verifying
geometry theorems by converting them to equations.  So far, it 
has not been made to work%
\footnote{Kenneth Manders has studied the role of diagrams in Euclidean proofs and argue that the use of diagrams in Euclid is limited to some class of properties~\cite{manders_euclidean_2011}. Luengo and Mumma have proposed formal systems which intend to capture the obvious spatial properties as built in inference rules~\cite{mumma_proofs_2010}, but both systems have been shown by Miller to be inconsistent~\cite{miller_diagrammatic_2001,miller_inconsistency_2012}. Miller has proposed another system, but the number of cases which should be considered renders the diagrammatic system difficult to use in practice~\cite{miller_euclid_2007}. 
}
, but even if it did work, we would still prefer the completely axiomatic
approach, which is part of the tradition extending all the way from 
Euclid, through Hilbert and Tarski, to the present.

Starting in 2007, and still continuing as this is written in 2018, 
the second author of this paper and Gabriel Braun 
have been busy computer-checking 
geometrical theorems in the proof assistant Coq.   
They verified Pappus's theorem \cite{narboux2017}
(which is important for the geometrical definition of arithmetic).
They verified that Hilbert's axioms follow from Tarski's 
\cite{narboux2012}. 
With Pierre Boutry, they verified that Tarski's axioms follow from Hilbert's \cite{narboux2016},
and completed \cite{narboux2016b} the verification of 
the theorems in Szmielew's part of \cite{schwabhauser}, which the 
second author began (with other co-authors) in \cite{narboux2015}.
Work is currently being done toward checking Euclid's propositions from Hilbert/Tarski axioms within Coq. This work differs from the work presented in this paper, because the goal is not to verify Euclid's \emph{proofs} but Euclid's \emph{statements} using an axiom system as minimal as possible.

For further information, please see the forthcoming
survey article \cite{NJFsurvey}. 
  
\section{How wrong was Euclid?}
The point that has given rise to the most discussion when our work has 
been presented is not whether our proofs are certifiably correct, but 
whether Euclid's proofs are really {\em wrong},  or  ``how wrong'' they
are.  Therefore we address this issue explicitly.   We classify the problems
with Euclid's proofs into 
\begin{itemize}
\item Missing axioms (circle axioms, Pasch, betweenness on a line)
\item Gaps (correctable failures to prove collinearity or non-collinearity or betweenness)
\item Superfluous axioms: theory of angles (angle equality and order can be defined and its properties proved, rather than assumed).  
\item Difficult theorems ``justified'' by common notions. For example, it 
can be proved that an angle cannot be less than itself. 
Euclid uses this at the end of I.7 without given any justification at all. 
Hilbert does no better: he assumes it by 
including uniqueness in his angle-copying axiom. 
\item Superfluous axioms:  Postulate IV (all right angles are equal) can be proved,  though a correct proof is rather difficult.
\item Out and out errors. We have mentioned Euclid's incorrect proof of 
the angle bisection proposition I.9, and his uncorrectable proof of I.7.
Euclid's errors, and our correct proofs of his propositions, deserve 
a full discussion, but that will be lengthy and is postponed to a 
future publication. Our purpose in this paper is to discuss
the proof-checking, not the geometry, i.e., not the proofs themselves.
\end{itemize}

\section{Conclusion}
Our aim was to remove every flaw from Euclid's axioms, definitions, postulates, and common notions,
and give formal proofs of all the propositions in Book I.  Did we achieve that aim?

The statements of the postulates and the definitions needed little if any change;  it is the 
axioms and proofs that needed corrections.   
We replaced Euclid's axioms and postulates by similar ones in a language similar to Tarski's,
but using strict betweenness.  We added line-circle and circle-circle axioms, and both 
inner and outer Pasch; we added the five-line axiom to enable a correct proof of the SAS
congruence criterion (Prop.~I.2).  We dropped Postulate 4 (all right angles are equal) because 
it can be proved,  and formulated Postulate 5 (the parallel postulate) in our points-only language.  
Inequality of lines and angles, and equality of angles, become defined concepts and the common notions concerning those 
concepts become theorems.  We used Tarski's definition of ``same side'', an essential concept which 
Euclid mentioned but neither defined nor considered as a common notion. 
We used Euclid's rather than Tarski's extension axiom, so that Euclid's
I.2 would not be superfluous.   We think that this 
choice of axioms is very close to Euclid's.   

With this choice of axioms, we were able  to prove Euclid's propositions I.1 to
I.48.
These proofs follow Euclid as closely as possible, and have been checked
in two well-known and respected proof checkers.  We have therefore shown beyond a shadow of a doubt that these proofs are correct.  Then we checked, again using those 
same proof checkers, that the axioms we used hold in $\R^2$.
 In particular all of Euclid's propositions in Book~I {\em and corrected proofs of those propositions, close to Euclid's ideas,} 
are valid, without a 
shadow of a doubt, in Euclidean two-space.%
\footnote{They hold in Euclidean
$n$-space too, since there is no dimension axiom, but we did
not formally check that.}
 While many paper-and-pencil
formalizations of 
Euclid have been put forward in the past, we are the first to be able to make this claim.
That this was not a superfluous exercise is shown by the many difficulties we encountered,
and the fact that we had to prove the propositions in quite a different order than Euclid,
and in some cases by different proofs.  In this paper we have focused on the axioms 
and the proof checking.  A subsequent publication will present the geometrical difficulties and compare our proofs in detail to those of Euclid. 

To play the devil's advocate, what argument could be made that we did not achieve the aim 
stated above?   One might complain that we proved the propositions in a different order than 
Euclid did.   We had to do that, because we could not prove them in the original order 
using our axioms.   The devil might argue that we should have strengthened the circle-circle
axiom to provide for an intersection point of the two circles on a given side of the line 
joining the centers.   With this stronger axiom we could have fixed Euclid's proof of I.9
(angle bisection) and used it as Euclid did to bisect a line.  But this would amount 
to assuming, rather than proving, the existence of erected perpendiculars, midpoints, and angle 
bisectors.  Besides, it would not have 
fixed the other problems we had with the ordering of theorems, and the proof we gave, using
midpoints to bisect angles rather than bisection of angles to construct midpoints,  is 
beautiful,  even if it was discovered by Gupta more than two thousand years after Euclid. 

We think that the devil would be wrong to say we should have strengthened circle-circle,
and we therefore claim that we did indeed
remove every flaw from Euclid's axioms, definitions, postulates, and common notions,
and give correct proofs of the propositions in Book I.

\section{Appendix 1: Formal proof of Prop.~I.1}

The reader may compare the following proof to Euclid's.  The conclusion
{\tt ELABC},  that $ABC$ is equilateral,  is reached about halfway through,
and that corresponds to the end of Euclid's proof.  The firsts half 
of our proof corresponds fairly naturally to Euclid's, except for 
quoting the circle-circle axiom, and verifying that is hypotheses are 
satisfied.

 The last half of our 
proof is devoted to proving that $ABC$ is a triangle, that is, the three
points are not collinear.  Note the use of lemma \goodbreak\noindent{\tt partnotequalwhole}.  If 
Euclid had noticed the need to prove that $ABC$ actually is a triangle,
he would have justified it using the common notions, applied to 
equality (congruence) of lines.  This version of ``the part is not 
equal to the whole'' is not an axiom for us, but a theorem.  

At the request of the referee we present this proof in a typeset form 
rather than in its native Polish form.  Obviously further mechanical 
processing can increase its superficial resemblance to Euclid's style,
but the point of our present work is simply its mechanically-checked
{\em correctness}. 

\def\text#1{\mbox{#1}}

\newcommand{\BetS}[3]{ #2 \text{ is strictly between } #1 \text{ and } #3  }
\newcommand{\eq}[2]{ { #1 = #2 }  }
\newcommand{\noteq}[2]{ { #1 \neq #2 }  }
\newcommand{\Col}[3] { #1 #2 #3 \text{ are collinear} }
\newcommand{\nCol}[3] { #1 #2 #3 \text{ are not collinear} }
\newcommand{\Cong}[4] { { #1 #2 \cong #3 #4 } }
\newcommand{\NotCong}[4] { { #1 #2 \not\cong #3 #4 } }
\newcommand{\Triangle}[3] { { #1 #2 #3 \text{ is a triangle} } }
\newcommand{\CongA}[6] { {\angle #1 #2 #3 \cong \angle #4 #5 #6 } }
\newcommand{\OnCirc}[2] { { #1 \text{ is on circle } #2 } }
\newcommand{\Out}[3] { #3 \text{ lies on ray } #1#2  }
\newcommand{\InCirc}[2] { #1 \text{ is inside circle } #2 }
\newcommand{\OutCirc}[2] { { #1 \text{ is outside circle } #2 } }
\newcommand{\CI}[4] { #1 \text{ is the circle of center } #2 \text{ and radius } #3#4 }
\newcommand{\Cut}[5] { #1#2 \text{ cut } #3#4 \text{ in } #5 }
\newcommand{\Supp}[5] { #1#2#3 \text{ and } #4#2#5 \text{ are supplementary angles}}
\newcommand{\TC}[6] { \text{Triangle } #1#2#3 \text{ is congruent to } #4#5#6 }
\newcommand{\Lt}[4] { {#1#2 < #3#4} }
\newcommand{\SameSide}[4] { #1#2 \text{ are on the same side of } #3#4 }
\newcommand{\OS}[4] { #1#4 \text{ are on opposite sides of } #2#3 }
\newcommand{\LtA}[6] {{\angle #1#2#3 < \angle #4#5#6 }}
\newcommand{\equilateral}[3] { #1#2#3 \text{ is equilateral }}
\newcommand{\Midpoint}[3] { #2 \text{ is the midpoint of } #1#3}
\newcommand{\isosceles}[3] { #1#2#3 \text{ is isosceles with base } #2#3}
\newcommand{\Perp}[4] { #1#2 \perp #3#4}
\newcommand{\TG}[6] { #1#2 \text{ and } #3#4 \text{ are together greater than } #5#6}
\newcommand{\InAngle}[4] { #4 \text{ is in the interior of angle } #1#2#3 }
\newcommand{\RT}[6] { #1#2#3 \text{ and } #4#5#6 \text{ make together two right angles} }
\newcommand{\SumA}[9] {#1#2#3 \text{ and } #4#5#6 \text{ are together equal to } #7#8#9}
\newcommand{\Meet}[4] { \text{ line } #1#2 \text{ crosses } #3#4 }
\newcommand{\EF}[8] {#1#2#3#4 \text{ and } #5#6#7#8 \text{ are equal quadrilaterals} }
\newcommand{\SQ}[4] { #1#2#3#4 \text{ is a square }}
\newcommand{\ET}[6] { #1#2#3 \text{ and } #4#5#6 \text{ are equal triangles}}
\newcommand{\TE}[6] { #1#2#3 \text{ and } #4#5#6 \text{ are equal triangles}}
\newcommand{\Par}[4] { {#1#2 \parallel #3#4 }}
\newcommand{\TP}[4] { #1#2 \text{ and } #3#4 \text{ are Tarski parallel }}
\newcommand{\RE}[4] { #1#2#3#4 \text{ is a rectangle} }
\newcommand{\ER}[8] { #1#2#3#4 \text{ and } #5#6#7#8 \text{ are equal rectangles} }
\newcommand{\RC}[8] { #1#2#3#4 \text{ and } #5#6#7#8 \text{ are congruent rectangles} }
\newcommand{\BR}[5] { #1#2#3#4 \text{ is a base rectangle of triangle } #2#3#5 }
\newcommand{\FE}[8] { #1#2#3#4 \text{ and } #5#6#7#8 \text{ are equal quadrilaterals}}
\newcommand{\FR}[8] { #5#6#7#8 \text{ is a figure rectangle of } #1#2#3#4  }
\newcommand{\TT}[8] { #1#2,#2#4 \text{ are together greater than } #5#6,#7#8}
\newcommand{\PG}[4] { #1#2#3#4 \text{ is a parallelogram }}
\newcommand{\CRR}[4] {#1#2 \text{ crosses } #3#4 }
\newcommand{\Per}[3] { #1#2#3 \text{ is a right angle}}
\newcommand{\PerpAt}[5] { #1#2 \text{ is perpendicular to } #3#4 \text{ at } #5 }

\def\myindentzero {}
\def\myindentone  {\hskip 0.5cm}

\section*{Appendix 1: Formal proof of Prop. I.1}

\begin{Proposition}[Prop. I.1]
\label{proposition-01}
On a given finite straight line to construct an equilateral triangle.
\begin{dmath*} 
 \forall A B\qquad 
   \noteq{A}{B} \implies 
   \exists X\qquad \equilateral{A}{B}{X} \land \Triangle{A}{B}{X}
\end{dmath*}
\end{Proposition}
{\em Proof}
\myindentzero$\text{Let } J\;\text{be such that }\CI{J}{A}{A}{B}$ by postulate Euclid3.

\myindentzero$\noteq{B}{A}$ by lemma inequalitysymmetric.

\myindentzero$\text{Let } K\;\text{be such that }\CI{K}{B}{B}{A}$ by postulate Euclid3.

\myindentzero$\text{Let } D\;\text{be such that }\BetS{B}{A}{D} \;\land\; \Cong{A}{D}{A}{B}$ by lemma localextension.

\myindentzero$\Cong{A}{D}{B}{A}$ by lemma congruenceflip.

\myindentzero$\Cong{B}{A}{B}{A}$ by common notion congruencereflexive.

\myindentzero$\OutCirc{D}{K}$ by definition of outside.

\myindentzero$\eq{B}{B}$ by common notion equalityreflexive.

\myindentzero$\InCirc{B}{K}$ by definition of inside.

\myindentzero$\Cong{A}{B}{A}{B}$ by common notion congruencereflexive.

\myindentzero$\OnCirc{B}{J}$ by definition of on.

\myindentzero$\OnCirc{D}{J}$ by definition of on.

\myindentzero$\eq{A}{A}$ by common notion equalityreflexive.

\myindentzero$\InCirc{A}{J}$ by definition of inside.

\myindentzero$\text{Let } C\;\text{be such that }\OnCirc{C}{K} \;\land\; \OnCirc{C}{J}$ by postulate circle-circle.

\myindentzero$\Cong{A}{C}{A}{B}$ by axiom circle-center-radius.

\myindentzero$\Cong{A}{B}{A}{C}$ by lemma congruencesymmetric.

\myindentzero$\Cong{B}{C}{B}{A}$ by axiom circle-center-radius.

\myindentzero$\Cong{B}{C}{A}{B}$ by lemma congruenceflip.

\myindentzero$\Cong{B}{C}{A}{C}$ by lemma congruencetransitive.

\myindentzero$\Cong{A}{B}{B}{C}$ by lemma congruencesymmetric.

\myindentzero$\Cong{A}{C}{C}{A}$ by common notion equalityreverse.

\myindentzero$\Cong{B}{C}{C}{A}$ by lemma congruencetransitive.

\myindentzero$\equilateral{A}{B}{C}$ by definition of equilateral.

\myindentzero$\noteq{B}{C}$ by axiom nocollapse.

\myindentzero$\noteq{C}{A}$ by axiom nocollapse.

\myindentzero Let show that $\BetS{A}{C}{B}$ does not hold by contradiction:

\myindentzero\{

\myindentone$\NotCong{A}{C}{A}{B}$ by lemma partnotequalwhole.

\myindentone$\Cong{C}{A}{A}{C}$ by common notion equalityreverse.

\myindentone$\Cong{C}{A}{A}{B}$ by lemma congruencetransitive.

\myindentone$\Cong{A}{C}{C}{A}$ by common notion equalityreverse.

\myindentone$\Cong{A}{C}{A}{B}$ by lemma congruencetransitive.

\myindentone We have a contradiction.

\myindentzero\}

\myindentzero Let show that $\BetS{A}{B}{C}$ does not hold by contradiction:

\myindentzero\{

\myindentone$\NotCong{A}{B}{A}{C}$ by lemma partnotequalwhole.

\myindentone$\Cong{A}{B}{C}{A}$ by lemma congruencetransitive.

\myindentone$\Cong{C}{A}{A}{C}$ by common notion equalityreverse.

\myindentone$\Cong{A}{B}{A}{C}$ by lemma congruencetransitive.

\myindentone We have a contradiction.

\myindentzero\}

\myindentzero Let show that $\BetS{B}{A}{C}$ does not hold by contradiction:

\myindentzero\{

\myindentone$\NotCong{B}{A}{B}{C}$ by lemma partnotequalwhole.

\myindentone$\Cong{B}{A}{A}{B}$ by common notion equalityreverse.

\myindentone$\Cong{B}{A}{B}{C}$ by lemma congruencetransitive.

\myindentone We have a contradiction.

\myindentzero\}

\myindentzero Let show that $\Col{A}{B}{C}$ does not hold by contradiction:

\myindentzero\{

\myindentone$\noteq{A}{C}$ by lemma inequalitysymmetric.

\myindentone$\eq{A}{B} \;\lor\; \eq{A}{C} \;\lor\; \eq{B}{C} \;\lor\; \BetS{B}{A}{C} \;\lor\; \BetS{A}{B}{C} \;\lor\; \BetS{A}{C}{B}$ by definition of collinear.

\myindentone We have a contradiction.

\myindentzero\}

\myindentzero$\Triangle{A}{B}{C}$ by definition of triangle.

\section*{Appendix 2: Axioms and Definitions}
The following formulas are presented in a format that 
can be cut and pasted, even from a pdf file.

\subsection*{Definitions}

{\em \texttt{A and B are distinct points}}
\begin{lstlisting}[breaklines,breakatwhitespace,basicstyle=\ttfamily\footnotesize]
NE(A,B) := 
  ~ EQ(A,B)

\end{lstlisting}
{\em \texttt{A, B, and C are collinear}}
\begin{lstlisting}[breaklines,breakatwhitespace,basicstyle=\ttfamily\footnotesize]
CO(A,B,C) := 
  EQ(A,B) \/ EQ(A,C) \/ EQ(B,C) \/ BE(B,A,C) \/ BE(A,B,C) \/ BE(A,C,B)

\end{lstlisting}
{\em \texttt{A, B, and C are not collinear}}
\begin{lstlisting}[breaklines,breakatwhitespace,basicstyle=\ttfamily\footnotesize]
NC(A,B,C) := 
  NE(A,B) /\ NE(A,C) /\ NE(B,C) /\ ~BE(A,B,C) /\ ~BE(A,C,B) /\ ~BE(B,A,C)

\end{lstlisting}
{\em \texttt{P is inside (some) circle J of center C and radius AB}}
\begin{lstlisting}[breaklines,breakatwhitespace,basicstyle=\ttfamily\footnotesize]
IC(P,J) := 
  exists X Y U V W,  CI(J,U,V,W) /\ (EQ(P,U) \/ BE(U,Y,X) /\ EE(U,X,V,W) /\ EE(U,P,U,Y))

\end{lstlisting}
{\em \texttt{P is outside (some) circle J of center U and radius VW}}
\begin{lstlisting}[breaklines,breakatwhitespace,basicstyle=\ttfamily\footnotesize]
OC(P,J) := 
  exists X U V W,  CI(J,U,V,W) /\ BE(U,X,P) /\ EE(U,X,V,W)

\end{lstlisting}
{\em \texttt{B is on (some) circle J of center U and radius XY}}
\begin{lstlisting}[breaklines,breakatwhitespace,basicstyle=\ttfamily\footnotesize]
ON(B,J) := 
  exists X Y U,  CI(J,U,X,Y) /\ EE(U,B,X,Y)

\end{lstlisting}
{\em \texttt{ABC is equilateral}}
\begin{lstlisting}[breaklines,breakatwhitespace,basicstyle=\ttfamily\footnotesize]
EL(A,B,C) := 
  EE(A,B,B,C) /\ EE(B,C,C,A)

\end{lstlisting}
{\em \texttt{ABC is a triangle}}
\begin{lstlisting}[breaklines,breakatwhitespace,basicstyle=\ttfamily\footnotesize]
TR(A,B,C) := 
  NC(A,B,C)

\end{lstlisting}
{\em \texttt{C lies on ray AB}}
\begin{lstlisting}[breaklines,breakatwhitespace,basicstyle=\ttfamily\footnotesize]
RA(A,B,C) := 
  exists X,  BE(X,A,C) /\ BE(X,A,B)

\end{lstlisting}
{\em \texttt{AB is less than CD}}
\begin{lstlisting}[breaklines,breakatwhitespace,basicstyle=\ttfamily\footnotesize]
LT(A,B,C,D) := 
  exists X,  BE(C,X,D) /\ EE(C,X,A,B)

\end{lstlisting}
{\em \texttt{B is the midpoint of AC}}
\begin{lstlisting}[breaklines,breakatwhitespace,basicstyle=\ttfamily\footnotesize]
MI(A,B,C) := 
  BE(A,B,C) /\ EE(A,B,B,C)

\end{lstlisting}
{\em \texttt{Angle ABC is equal to angle abc}}
\begin{lstlisting}[breaklines,breakatwhitespace,basicstyle=\ttfamily\footnotesize]
EA(A,B,C,a,b,c) := 
  exists U V u v,  RA(B,A,U) /\ RA(B,C,V) /\ RA(b,a,u) /\ RA(b,c,v) /\ EE(B,U,b,u) /\ EE(B,V,b,v) /\ EE(U,V,u,v) /\ NC(A,B,C)

\end{lstlisting}
{\em \texttt{DBF is a supplement of ABC}}
\begin{lstlisting}[breaklines,breakatwhitespace,basicstyle=\ttfamily\footnotesize]
SU(A,B,C,D,F) := 
  RA(B,C,D) /\ BE(A,B,F)

\end{lstlisting}
{\em \texttt{ABC is a right angle}}
\begin{lstlisting}[breaklines,breakatwhitespace,basicstyle=\ttfamily\footnotesize]
RR(A,B,C) := 
  exists X,  BE(A,B,X) /\ EE(A,B,X,B) /\ EE(A,C,X,C) /\ NE(B,C)

\end{lstlisting}
{\em \texttt{PQ is perpendicular to AB at C and NCABP}}
\begin{lstlisting}[breaklines,breakatwhitespace,basicstyle=\ttfamily\footnotesize]
PA(P,Q,A,B,C) := 
  exists X,  CO(P,Q,C) /\ CO(A,B,C) /\ CO(A,B,X) /\ RR(X,C,P)

\end{lstlisting}
{\em \texttt{PQ is perpendicular to AB}}
\begin{lstlisting}[breaklines,breakatwhitespace,basicstyle=\ttfamily\footnotesize]
PE(P,Q,A,B) := 
  exists X,  PA(P,Q,A,B,X)

\end{lstlisting}
{\em \texttt{P is in the interior of angle ABC}}
\begin{lstlisting}[breaklines,breakatwhitespace,basicstyle=\ttfamily\footnotesize]
IA(A,B,C,P) := 
  exists X Y,  RA(B,A,X) /\ RA(B,C,Y) /\ BE(X,P,Y)

\end{lstlisting}
{\em \texttt{P and Q are on opposite sides of AB}}
\begin{lstlisting}[breaklines,breakatwhitespace,basicstyle=\ttfamily\footnotesize]
OS(P,A,B,Q) := 
  exists X,  BE(P,X,Q) /\ CO(A,B,X) /\ NC(A,B,P)

\end{lstlisting}
{\em \texttt{P and Q are on the same side of AB}}
\begin{lstlisting}[breaklines,breakatwhitespace,basicstyle=\ttfamily\footnotesize]
SS(P,Q,A,B) := 
  exists X U V,  CO(A,B,U) /\ CO(A,B,V) /\ BE(P,U,X) /\ BE(Q,V,X) /\ NC(A,B,P) /\ NC(A,B,Q)

\end{lstlisting}
{\em \texttt{ABC is isosceles with base BC}}
\begin{lstlisting}[breaklines,breakatwhitespace,basicstyle=\ttfamily\footnotesize]
IS(A,B,C) := 
  TR(A,B,C) /\ EE(A,B,A,C)

\end{lstlisting}
{\em \texttt{AB cuts CD in E}}
\begin{lstlisting}[breaklines,breakatwhitespace,basicstyle=\ttfamily\footnotesize]
CU(A,B,C,D,E) := 
  BE(A,E,B) /\ BE(C,E,D) /\ NC(A,B,C) /\ NC(A,B,D)

\end{lstlisting}
{\em \texttt{Triangle ABC is congruent to abc}}
\begin{lstlisting}[breaklines,breakatwhitespace,basicstyle=\ttfamily\footnotesize]
TC(A,B,C,a,b,c) := 
  EE(A,B,a,b) /\ EE(B,C,b,c) /\ EE(A,C,a,c) /\ TR(A,B,C)

\end{lstlisting}
{\em \texttt{Angle ABC is less than angle DEF}}
\begin{lstlisting}[breaklines,breakatwhitespace,basicstyle=\ttfamily\footnotesize]
AO(A,B,C,D,E,F) := 
  exists U X V,  BE(U,X,V) /\ RA(E,D,U) /\ RA(E,F,V) /\ EA(A,B,C,D,E,X)

\end{lstlisting}
{\em \texttt{AB and CD are together greater than EF}}
\begin{lstlisting}[breaklines,breakatwhitespace,basicstyle=\ttfamily\footnotesize]
TG(A,B,C,D,E,F) := 
  exists X,  BE(A,B,X) /\ EE(B,X,C,D) /\ LT(E,F,A,X)

\end{lstlisting}
{\em \texttt{AB, CD are together greater than EF,GH}}
\begin{lstlisting}[breaklines,breakatwhitespace,basicstyle=\ttfamily\footnotesize]
TT(A,B,C,D,E,F,G,H) := 
  exists X,  BE(E,F,X) /\ EE(F,X,G,H) /\ TG(A,B,C,D,E,X)

\end{lstlisting}
{\em \texttt{ABC and DEF make together two right angles}}
\begin{lstlisting}[breaklines,breakatwhitespace,basicstyle=\ttfamily\footnotesize]
RT(A,B,C,D,E,F) := 
  exists X Y Z U V,  SU(X,Y,U,V,Z) /\ EA(A,B,C,X,Y,U) /\ EA(D,E,F,V,Y,Z)

\end{lstlisting}
{\em \texttt{AB meets CD}}
\begin{lstlisting}[breaklines,breakatwhitespace,basicstyle=\ttfamily\footnotesize]
ME(A,B,C,D) := 
  exists X,  NE(A,B) /\ NE(C,D) /\ CO(A,B,X) /\ CO(C,D,X)

\end{lstlisting}
{\em \texttt{AB crosses CD}}
\begin{lstlisting}[breaklines,breakatwhitespace,basicstyle=\ttfamily\footnotesize]
CR(A,B,C,D) := 
  exists X,  BE(A,X,B) /\ BE(C,X,D)

\end{lstlisting}
{\em \texttt{AB and CD are Tarski parallel}}
\begin{lstlisting}[breaklines,breakatwhitespace,basicstyle=\ttfamily\footnotesize]
TP(A,B,C,D) := 
  NE(A,B) /\ NE(C,D) /\ ~ ME(A,B,C,D) /\ SS(C,D,A,B)

\end{lstlisting}
{\em \texttt{AB and CD are parallel}}
\begin{lstlisting}[breaklines,breakatwhitespace,basicstyle=\ttfamily\footnotesize]
PR(A,B,C,D) := 
  exists U V u v X,  NE(A,B) /\ NE(C,D) /\ CO(A,B,U) /\ CO(A,B,V) /\ NE(U,V) /\ CO(C,D,u) /\ CO(C,D,v) /\ NE(u,v) /\ ~ ME(A,B,C,D) /\ BE(U,X,v) /\ BE(u,X,V)

\end{lstlisting}
{\em \texttt{ABC and DEF are together equal to PQR}}
\begin{lstlisting}[breaklines,breakatwhitespace,basicstyle=\ttfamily\footnotesize]
AS(A,B,C,D,E,F,P,Q,R) := 
  exists X,  EA(A,B,C,P,Q,X) /\ EA(D,E,F,X,Q,R) /\ BE(P,X,R)

\end{lstlisting}
{\em \texttt{ABCD is a parallelogram}}
\begin{lstlisting}[breaklines,breakatwhitespace,basicstyle=\ttfamily\footnotesize]
PG(A,B,C,D) := 
  PR(A,B,C,D) /\ PR(A,D,B,C)

\end{lstlisting}
{\em \texttt{ABCD is a square}}
\begin{lstlisting}[breaklines,breakatwhitespace,basicstyle=\ttfamily\footnotesize]
SQ(A,B,C,D) := 
  EE(A,B,C,D) /\ EE(A,B,B,C) /\ EE(A,B,D,A) /\ RR(D,A,B) /\ RR(A,B,C) /\ RR(B,C,D) /\ RR(C,D,A)

\end{lstlisting}
{\em \texttt{ABCD is a rectangle}}
\begin{lstlisting}[breaklines,breakatwhitespace,basicstyle=\ttfamily\footnotesize]
RE(A,B,C,D) := 
  RR(D,A,B) /\ RR(A,B,C) /\ RR(B,C,D) /\ RR(C,D,A) /\ CR(A,C,B,D)

\end{lstlisting}
{\em \texttt{ABCD and abcd are congruent rectangles}}
\begin{lstlisting}[breaklines,breakatwhitespace,basicstyle=\ttfamily\footnotesize]
RC(A,B,C,D,a,b,c,d) := 
  RE(A,B,C,D) /\ RE(a,b,c,d) /\ EE(A,B,a,b) /\ EE(B,C,b,c)

\end{lstlisting}
{\em \texttt{ABCD and abcd are equal rectangles}}
\begin{lstlisting}[breaklines,breakatwhitespace,basicstyle=\ttfamily\footnotesize]
ER(A,B,C,D,a,b,c,d) := 
  exists X Y Z U x z u w W,  RC(A,B,C,D,X,Y,Z,U) /\ RC(a,b,c,d,x,Y,z,u) /\ BE(x,Y,Z) /\ BE(X,Y,z) /\ BE(W,U,w)

\end{lstlisting}
{\em \texttt{ABCD is a base rectangle of triangle BCE}}
\begin{lstlisting}[breaklines,breakatwhitespace,basicstyle=\ttfamily\footnotesize]
BR(A,B,C,D,E) := 
  RE(B,C,D,E) /\ CO(D,E,A)

\end{lstlisting}
{\em \texttt{ABC and abc are equal triangles}}
\begin{lstlisting}[breaklines,breakatwhitespace,basicstyle=\ttfamily\footnotesize]
TE(A,B,C,a,b,c) := 
  exists X Y x y,  RE(A,B,X,Y) /\ RE(a,b,x,y) /\ CO(X,Y,C) /\ CO(x,y,c) /\ ER(A,B,X,Y,a,b,x,y)

\end{lstlisting}
{\em \texttt{ABCD and abcd are equal quadrilaterals}}
\begin{lstlisting}[breaklines,breakatwhitespace,basicstyle=\ttfamily\footnotesize]
FE(A,B,C,D,a,b,c,d) := 
  exists X Y Z U x y z u,  OS(A,B,C,D) /\ OS(a,b,c,d) /\ FR(A,B,C,D,X,Y,Z,U) /\ FR(a,b,c,d,x,y,z,u) /\ ER(X,Y,Z,U,x,y,z,u)

\end{lstlisting}

\subsection*{Common Notions}

\begin{lstlisting}[breaklines,breakatwhitespace,basicstyle=\ttfamily\footnotesize]
cn-equalitytransitive
forall A B C, EQ(A,C) /\ EQ(B,C) ==> EQ(A,B)\end{lstlisting}
\begin{lstlisting}[breaklines,breakatwhitespace,basicstyle=\ttfamily\footnotesize]
cn-congruencetransitive
forall B C D E P Q, EE(P,Q,B,C) /\ EE(P,Q,D,E) ==> EE(B,C,D,E)\end{lstlisting}
\begin{lstlisting}[breaklines,breakatwhitespace,basicstyle=\ttfamily\footnotesize]
cn-equalityreflexive
forall A, EQ(A,A)\end{lstlisting}
\begin{lstlisting}[breaklines,breakatwhitespace,basicstyle=\ttfamily\footnotesize]
cn-congruencereflexive
forall A B, EE(A,B,A,B)\end{lstlisting}
\begin{lstlisting}[breaklines,breakatwhitespace,basicstyle=\ttfamily\footnotesize]
cn-equalityreverse
forall A B, EE(A,B,B,A)\end{lstlisting}
\begin{lstlisting}[breaklines,breakatwhitespace,basicstyle=\ttfamily\footnotesize]
cn-sumofparts
forall A B C a b c, EE(A,B,a,b) /\ EE(B,C,b,c) /\ BE(A,B,C) /\ BE(a,b,c) ==> EE(A,C,a,c)\end{lstlisting}
\begin{lstlisting}[breaklines,breakatwhitespace,basicstyle=\ttfamily\footnotesize]
cn-stability
forall A B, ~ NE(A,B) ==> EQ(A,B)\end{lstlisting}
\begin{lstlisting}[breaklines,breakatwhitespace,basicstyle=\ttfamily\footnotesize]
cn-equalitysub
forall A B C D, EQ(D,A) /\ BE(A,B,C) ==> BE(D,B,C)\end{lstlisting}

\subsection*{Axioms of betweenness and congruence}

\begin{lstlisting}[breaklines,breakatwhitespace,basicstyle=\ttfamily\footnotesize]
axiom-betweennessidentity
forall A B, ~BE(A,B,A)\end{lstlisting}
\begin{lstlisting}[breaklines,breakatwhitespace,basicstyle=\ttfamily\footnotesize]
axiom-betweennesssymmetry
forall A B C, BE(A,B,C) ==> BE(C,B,A)\end{lstlisting}
\begin{lstlisting}[breaklines,breakatwhitespace,basicstyle=\ttfamily\footnotesize]
axiom-innertransitivity
forall A B C D, BE(A,B,D) /\ BE(B,C,D) ==> BE(A,B,C)\end{lstlisting}
\begin{lstlisting}[breaklines,breakatwhitespace,basicstyle=\ttfamily\footnotesize]
axiom-connectivity
forall A B C D, BE(A,B,D) /\ BE(A,C,D) /\ ~BE(A,B,C) /\ ~BE(A,C,B) ==> EQ(B,C)\end{lstlisting}
\begin{lstlisting}[breaklines,breakatwhitespace,basicstyle=\ttfamily\footnotesize]
axiom-nocollapse
forall A B C D, NE(A,B) /\ EE(A,B,C,D) ==> NE(C,D)\end{lstlisting}
\begin{lstlisting}[breaklines,breakatwhitespace,basicstyle=\ttfamily\footnotesize]
axiom-5-line
forall A B C D a b c d, EE(B,C,b,c) /\ EE(A,D,a,d) /\ EE(B,D,b,d) /\ BE(A,B,C) /\ BE(a,b,c) /\ EE(A,B,a,b) ==> EE(D,C,d,c)\end{lstlisting}

\subsection*{Postulates}

\begin{lstlisting}[breaklines,breakatwhitespace,basicstyle=\ttfamily\footnotesize]
postulate-Pasch-inner
forall A B C P Q, BE(A,P,C) /\ BE(B,Q,C) /\ NC(A,C,B) ==> exists X, BE(A,X,Q) /\ BE(B,X,P)\end{lstlisting}
\begin{lstlisting}[breaklines,breakatwhitespace,basicstyle=\ttfamily\footnotesize]
postulate-Pasch-outer
forall A B C P Q, BE(A,P,C) /\ BE(B,C,Q) /\ NC(B,Q,A) ==> exists X, BE(A,X,Q) /\ BE(B,P,X)\end{lstlisting}
\begin{lstlisting}[breaklines,breakatwhitespace,basicstyle=\ttfamily\footnotesize]
postulate-Euclid2
forall A B, NE(A,B) ==> exists X, BE(A,B,X)\end{lstlisting}
\begin{lstlisting}[breaklines,breakatwhitespace,basicstyle=\ttfamily\footnotesize]
postulate-Euclid3
forall A B, NE(A,B) ==> exists X, CI(X,A,A,B)\end{lstlisting}
\begin{lstlisting}[breaklines,breakatwhitespace,basicstyle=\ttfamily\footnotesize]
postulate-line-circle
forall A B C K P Q, CI(K,C,P,Q) /\ IC(B,K) /\ NE(A,B) ==> exists X Y, CO(A,B,X) /\ BE(A,B,Y) /\ ON(X,K) /\ ON(Y,K) /\ BE(X,B,Y)\end{lstlisting}
\begin{lstlisting}[breaklines,breakatwhitespace,basicstyle=\ttfamily\footnotesize]
postulate-circle-circle
forall C D F G J K P Q R S, CI(J,C,R,S) /\ IC(P,J) /\ OC(Q,J) /\ CI(K,D,F,G) /\ ON(P,K) /\ ON(Q,K) ==> exists X, ON(X,J) /\ ON(X,K)\end{lstlisting}
\begin{lstlisting}[breaklines,breakatwhitespace,basicstyle=\ttfamily\footnotesize]
postulate-Euclid5
forall a p q r s t, BE(r,t,s) /\ BE(p,t,q) /\ BE(r,a,q) /\ EE(p,t,q,t) /\ EE(t,r,t,s) /\ NC(p,q,s) ==> exists X, BE(p,a,X) /\ BE(s,q,X)\end{lstlisting}
\begin{lstlisting}[breaklines,breakatwhitespace,basicstyle=\ttfamily\footnotesize]
axiom-circle-center-radius
forall A B C J P, CI(J,A,B,C) /\ ON(P,J) ==> EE(A,P,B,C)\end{lstlisting}

\subsection*{Axioms for Equal Figures}

\begin{lstlisting}[breaklines,breakatwhitespace,basicstyle=\ttfamily\footnotesize]
axiom-congruentequal
forall A B C a b c, TC(A,B,C,a,b,c) ==> ET(A,B,C,a,b,c)\end{lstlisting}
\begin{lstlisting}[breaklines,breakatwhitespace,basicstyle=\ttfamily\footnotesize]
axiom-ETpermutation
forall A B C a b c, ET(A,B,C,a,b,c) ==> ET(A,B,C,b,c,a) /\ ET(A,B,C,a,c,b) /\ ET(A,B,C,b,a,c) /\ ET(A,B,C,c,b,a) /\ ET(A,B,C,c,a,b)\end{lstlisting}
\begin{lstlisting}[breaklines,breakatwhitespace,basicstyle=\ttfamily\footnotesize]
axiom-ETsymmetric
forall A B C a b c, ET(A,B,C,a,b,c) ==> ET(a,b,c,A,B,C)\end{lstlisting}
\begin{lstlisting}[breaklines,breakatwhitespace,basicstyle=\ttfamily\footnotesize]
axiom-EFpermutation
forall A B C D a b c d, EF(A,B,C,D,a,b,c,d) ==> EF(A,B,C,D,b,c,d,a) /\ EF(A,B,C,D,d,c,b,a) /\ EF(A,B,C,D,c,d,a,b) /\ EF(A,B,C,D,b,a,d,c) /\ EF(A,B,C,D,d,a,b,c) /\ EF(A,B,C,D,c,b,a,d) /\ EF(A,B,C,D,a,d,c,b)\end{lstlisting}
\begin{lstlisting}[breaklines,breakatwhitespace,basicstyle=\ttfamily\footnotesize]
axiom-halvesofequals
forall A B C D a b c d, ET(A,B,C,B,C,D) /\ OS(A,B,C,D) /\ ET(a,b,c,b,c,d) /\ OS(a,b,c,d) /\ EF(A,B,D,C,a,b,d,c) ==> ET(A,B,C,a,b,c)\end{lstlisting}
\begin{lstlisting}[breaklines,breakatwhitespace,basicstyle=\ttfamily\footnotesize]
axiom-EFsymmetric
forall A B C D a b c d, EF(A,B,C,D,a,b,c,d) ==> EF(a,b,c,d,A,B,C,D)\end{lstlisting}
\begin{lstlisting}[breaklines,breakatwhitespace,basicstyle=\ttfamily\footnotesize]
axiom-EFtransitive
forall A B C D P Q R S a b c d, EF(A,B,C,D,a,b,c,d) /\ EF(a,b,c,d,P,Q,R,S) ==> EF(A,B,C,D,P,Q,R,S)\end{lstlisting}
\begin{lstlisting}[breaklines,breakatwhitespace,basicstyle=\ttfamily\footnotesize]
axiom-ETtransitive
forall A B C P Q R a b c, ET(A,B,C,a,b,c) /\ ET(a,b,c,P,Q,R) ==> ET(A,B,C,P,Q,R)\end{lstlisting}
\begin{lstlisting}[breaklines,breakatwhitespace,basicstyle=\ttfamily\footnotesize]
axiom-cutoff1
forall A B C D E a b c d e, BE(A,B,C) /\ BE(a,b,c) /\ BE(E,D,C) /\ BE(e,d,c) /\ ET(B,C,D,b,c,d) /\ ET(A,C,E,a,c,e) ==> EF(A,B,D,E,a,b,d,e)\end{lstlisting}
\begin{lstlisting}[breaklines,breakatwhitespace,basicstyle=\ttfamily\footnotesize]
axiom-cutoff2
forall A B C D E a b c d e, BE(B,C,D) /\ BE(b,c,d) /\ ET(C,D,E,c,d,e) /\ EF(A,B,D,E,a,b,d,e) ==> EF(A,B,C,E,a,b,c,e)\end{lstlisting}
\begin{lstlisting}[breaklines,breakatwhitespace,basicstyle=\ttfamily\footnotesize]
axiom-paste1
forall A B C D E a b c d e, BE(A,B,C) /\ BE(a,b,c) /\ BE(E,D,C) /\ BE(e,d,c) /\ ET(B,C,D,b,c,d) /\ EF(A,B,D,E,a,b,d,e) ==> ET(A,C,E,a,c,e)\end{lstlisting}
\begin{lstlisting}[breaklines,breakatwhitespace,basicstyle=\ttfamily\footnotesize]
axiom-deZolt1
forall B C D E, BE(B,E,D) ==> ~ ET(D,B,C,E,B,C)\end{lstlisting}
\begin{lstlisting}[breaklines,breakatwhitespace,basicstyle=\ttfamily\footnotesize]
axiom-deZolt2
forall A B C E F, TR(A,B,C) /\ BE(B,E,A) /\ BE(B,F,C) ==> ~ ET(A,B,C,E,B,F)\end{lstlisting}
\begin{lstlisting}[breaklines,breakatwhitespace,basicstyle=\ttfamily\footnotesize]
axiom-paste2
forall A B C D E M a b c d e m, BE(B,C,D) /\ BE(b,c,d) /\ ET(C,D,E,c,d,e) /\ EF(A,B,C,E,a,b,c,e) /\ BE(A,M,D) /\ BE(B,M,E) /\ BE(a,m,d) /\ BE(b,m,e) ==> EF(A,B,D,E,a,b,d,e)\end{lstlisting}
\begin{lstlisting}[breaklines,breakatwhitespace,basicstyle=\ttfamily\footnotesize]
axiom-paste3
forall A B C D M a b c d m, ET(A,B,C,a,b,c) /\ ET(A,B,D,a,b,d) /\ BE(C,M,D) /\ BE(A,M,B) \/ EQ(A,M) \/ EQ(M,B) /\ BE(c,m,d) /\ BE(a,m,b) \/ EQ(a,m) \/ EQ(m,b) ==> EF(A,C,B,D,a,c,b,d)\end{lstlisting}
\begin{lstlisting}[breaklines,breakatwhitespace,basicstyle=\ttfamily\footnotesize]
axiom-paste4
forall A B C D F G H J K L M P e m, EF(A,B,m,D,F,K,H,G) /\ EF(D,B,e,C,G,H,M,L) /\ BE(A,P,C) /\ BE(B,P,D) /\ BE(K,H,M) /\ BE(F,G,L) /\ BE(B,m,D) /\ BE(B,e,C) /\ BE(F,J,M) /\ BE(K,J,L) ==> EF(A,B,C,D,F,K,M,L)\end{lstlisting}

\bibliographystyle{siam}

\end{document}